\documentstyle[twocolumn,eqsecnum,prl,aps,epsfig,floats]{revtex}  %two columns
\newcommand{\Reop}{\mathop{\rm Re}\nolimits}
\newcommand{\Imop}{\mathop{\rm Im}\nolimits}

\newcommand{\Tr}{\mathop{\rm Tr}\nolimits}
\newcommand{\cbcex}{\langle\overline{\chi}\chi\rangle}

\newcommand{\chup}{$\chi U\phi$}
\newcommand{\eqref}[1]{(\ref{#1})}

\newcommand\fdfig[1]{%
  \psfig{file=#1,angle=90,width=\hsize,bbllx=65,bblly=15,bburx=540,bbury=780}}%
\newcommand\fdiifig[1]{%
  \psfig{file=#1,angle=90,width=\hsize,bbllx=65,bblly=15,bburx=540,bbury=780}}%
\begin{document}
%%%%%%%%%%%%%%%%%%%%%%%%%%%%%%%%%%%%%%%%%%%%%%%%%%%%%%%%%%%%%%%%%%%%%%%%%%%
\draft
\title{\hfill {\small PITHA 97/43, HLRZ1997\_66}\\
       Dynamical fermion mass generation \\
       at a tricritical point \\
       in strongly coupled U(1) lattice gauge theory 
%  \\  \underline{DRAFT 18.11.97}
}
\author{W.~Franzki and J.~Jers{\'a}k}
\address{Institut f{\"u}r
Theoretische Physik E, RWTH Aachen, Germany}
\date{\today}
\maketitle
%%%%%%%%%%%%%%%%%%%%%%%%%%%%%%%%%%%%%%%%%%%%%%%%%%%%%%%%%%%%%%%%%%%%%%
\begin{abstract}
  Fermion mass generation in the strongly coupled U(1) lattice gauge
  theory with fermion and scalar fields of equal charge is
  investigated by means of numerical simulation with dynamical
  fermions.  Chiral symmetry of this model is broken by the gauge
  interaction and restored by the light scalar. We present evidence
  for the existence of a particular, tricritical point of the
  corresponding phase boundary where the continuum limit might
  possibly be constructed. It is of interest as a model for dynamical
  symmetry breaking and mass generation due to a strong gauge
  interaction. In addition to the massive and unconfined fermion F and
  Goldstone boson $\pi$, a gauge ball of mass $m_S \simeq 1/2 m_F$ and
  some other states are found.  Tricritical exponents appear to be
  non-classical.

\end{abstract}
\narrowtext

%@@@@@@@@@@@@@@@@@@@@@@@@@@@@@@@@@@@@@@@@@@@@@@@@@@@@@@@@@@@@@@@@@@@@@@@
\section{Introduction}
%@@@@@@@@@@@@@@@@@@@@@@@@@@@@@@@@@@@@@@@@@@@@@@@@@@@@@@@@@@@@@@@@@@@@@@@

Attempts to construct a theory with dynamical breaking of global
chiral symmetries in four dimensions, which could explain or replace
the Higgs-Yukawa mechanism of particle mass generation, usually lead
to the introduction of a new strong gauge interaction beyond the
standard model and its standard extensions. For example the heavy top
quark and the idea of top condensate \cite{Na89} inspired the strongly
coupled topcolor and similar gauge models \cite{Ho87} (for a recent
overview see e.\,g.  \cite{Ko97}). Among the requirements such a
theory should satisfy, the most general ones are the following two:
First, because gauge theories tend to confine charges in a regime
where they break chiral symmetries dynamically, the physical states,
in particular fermions, must be composite singlets of the new gauge
symmetry.  Second, as a strong coupling regime is encountered, the
models should be nonperturbatively renormalizable in order to be
physically sensible in a sufficiently large interval of scales.

Even in very simplified models, this are too difficult dynamical problems to
get reliably under control by analytic means only.  Therefore, a numerical
investigation on the lattice of some prototypes of field theories with
the above properties may be instructive. In such an approach, the presumably
chiral character of the new gauge interaction and numerous phenomenological
aspects have to be left out of consideration. 

A promising candidate for such a prototype field theory on the lattice, the
$\chi U\phi$ model, has been described in Ref.~\cite{FrJe95a}.  Here the
four-dimensional vector-like U(1) gauge theory contains the staggered fermion
field $\chi$ and the scalar field $\phi$, both of unit charge. A Yukawa
coupling between these matter fields is prohibited by the gauge symmetry. The
global U(1) chiral symmetry, present when the bare mass $m_0$ of the fermion
field $\chi$ vanishes, is broken dynamically at strong gauge coupling $g$ by
the gauge interaction, similar to QCD or strongly coupled lattice QED.
Whereas both $\chi$ and $\phi$ constituents are confined, the massive physical
fermion $F = \phi^\dagger\chi$ with shielded charge appears.

The scalar {\em suppresses} the symmetry breaking when it gets lighter
and induces a phase transition to the chiral symmetric phase
\cite{LeShr87a,LeShi86d}.  At this transition, for large enough gauge
coupling, the mass of the physical fermion in lattice units scales,
$am_F \rightarrow 0$, and the lattice cutoff $1/a$ thus can be removed
for $m_F$ fixed in physical units. If the theory were renormalizable,
a continuum theory with massive fermion $F$, as well as a massless
Goldstone boson (``pion'' $\pi$) would be obtained.  When the global
U(1) chiral symmetry, modelling the SU(2) symmetry of the standard
model, is gauged, this $\pi$ boson is ``eaten'' by the corresponding
massive gauge boson. This is what is achieved in standard approaches
by the Higgs-Yukawa mechanism.

In this paper we address the question of renormalizability of the
\chup\ model at the line of chiral phase transitions induced by the
scalar field. We have no definite answer, but our extensive numerical
study of the model in the relevant region of the three-dimensional
parameter space (see Fig.~\ref{fig:pd4d3}) with dynamical fermions
provided several encouraging results:

%FFFFFFFFFFFFFFFFFFF Fig. 1 FFFFFFFFFFFFFFFFFFFFFFFFFFFFFFFFFFFFFFFFFF
\begin{figure*}
  \begin{center}
    \psfig{file=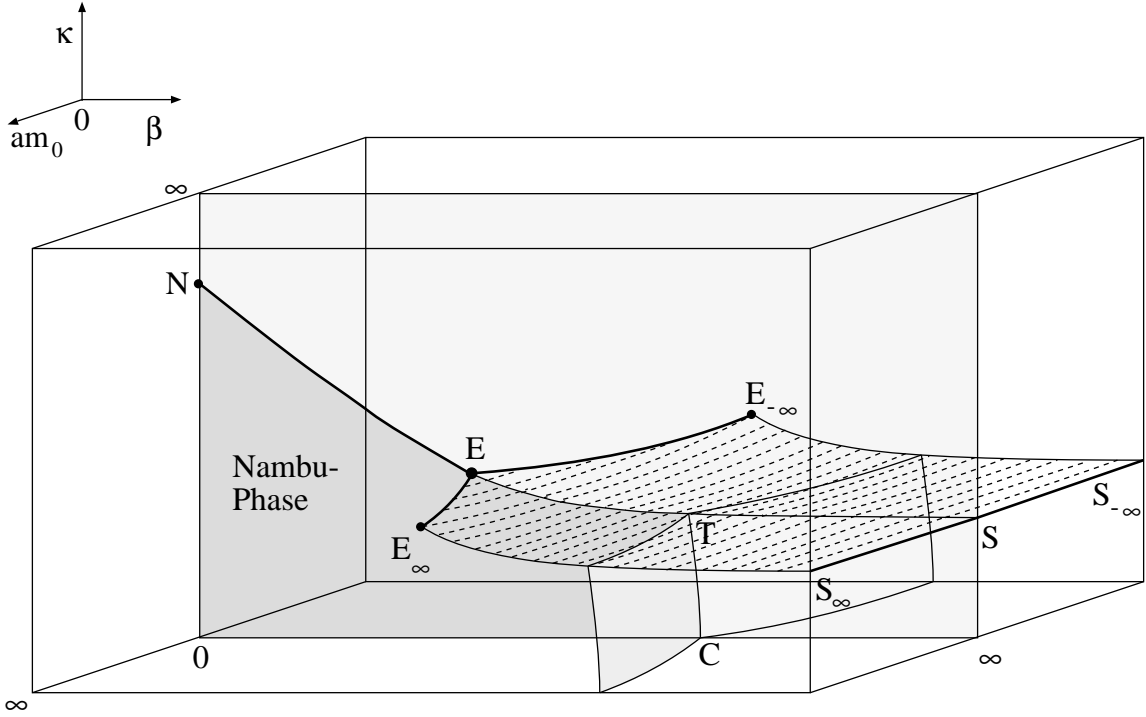,width=.85\hsize}%
    \caption[xxx]{%
      Schematic phase diagram of the \chup\ model. Three critical lines, NE,
      E$_\infty$E and E$_{-\infty}$E meet at the tricritical point E. The line
      NE is a part of the boundary of the Nambu phase (dark shaded region) at
      $m_0=0$ with spontaneously broken chiral symmetry, $am_F \ne 0$. This
      phase is a sheet of first order phase transitions at which the chiral
      condensate changes sign. The lines E$_\infty$E and E$_{-\infty}$E are
      critical boundaries of the ``wings'' of first order Higgs phase
      transitions. The light shaded region at $m_0=0$ corresponds to vanishing
      fermion mass, $am_F=0$. The vertical sheets containing the points T and
      C separate the confinement and Coulomb phases. The line ETS is a line of
      tripple points.}
    \label{fig:pd4d3}
  \end{center}
\end{figure*}%

\begin{enumerate}
  
\item Our previous studies \cite{FrFr95,FrFr96} have indicated that on the
  nearly whole chiral phase transition line, starting at the strong coupling
  limit $\beta = g^{-2} = 0$, the model behaves like the Nambu--Jona-Lasinio
  (NJL) model, belonging presumably to the same universality class. We now
  present strong evidence that at $\beta = \beta_E \simeq 0.62$ the line
  contains a special point, the tricritical point E, where for theoretical
  reasons the scaling behavior is different from the rest of the line. It is
  governed by another fixed point. In difference to the NJL model the gauge
  field is not auxiliary but plays an important dynamical role at the point E.

\item Using advanced methods of the finite size scaling analysis we estimate
  several tricritical exponents determining the scaling behavior at the point
  E and find nonclassical values. This indicates that due to the strong
  gauge interaction this point differs from the standard expectations for
  tricritical points in four dimensions \cite{LaSa84}.
  
\item In the vicinity of the point E in the broken phase, not only the fermion
  mass $am_F$, but also the masses of several bosons (neutral states composed
  of scalar and gauge fields) and mesons ($\overline{\chi} \chi$ states)
  scale, i.\,e.\ in lattice units they approach zero with constant ratios.
  This suggests a rich spectrum if the continuum limit of the model is
  approached at the point E. In particular, a gauge ball of mass $m_S \simeq
  1/2\, m_F$ is observed.  
  
\item The composite Goldstone state $\pi$ with properties required by chiral
  symmetry breaking is present.
  
\item We determine the effective Yukawa coupling $y_{\rm R}$ between the
  composite $F$ and $\pi$ states in the vicinity of the critical line and find
  that lines of constant $y_{\rm R}$ tend to approach the point E
  \cite{FrJe98a}. We cannot yet say whether some of them end at this point,
  which would imply a nontrivial continuum limit. However, this approaching
  means that the coupling decreases only slowly with an increasing cutoff
  $1/a$ on paths towards E, thus increasing the chances for renormalizability.

\end{enumerate}

We have not been able to achieve at least qualitative results in two
issues of major interest: A heavy scalar $\sigma$-meson, which would
correspond to a composite Higgs boson, is seen, but its mass in
lattice units does not yet scale on the lattices of sizes we could
afford and is strongly dependent on the bare fermion mass $am_0$. We
cannot say anything about its value in the continuum limit at E. Also
the pion decay constant $f_\pi$ does not scale, i.e. $f_\pi/m_F$ seems
to increase with decreasing distance from E. Its current value (at
$am_0 \simeq 0.4$) is about 1/3. The present data are consistent both
with the possibility that $f_\pi$ diverges in physical units, which
would indicate triviality \cite{HaKo97}, and that the absence of
scaling is due to too small lattices.

Concerning the possible triviality, we point out again \cite{FrJe95a} that the
$\chi U\phi$ model would be a valuable model even if the cutoff cannot be
removed completely without loosing the interaction, provided the cutoff
dependence of the renormalized couplings is sufficiently weak, e.g.
logarithmic as in the standard model. A dynamical approach to the Higgs-Yukawa
mechanism does not necessarily require a nontrivial fixed point. The
Higgs-Yukawa sector, whose validity is restricted due to the triviality by a
certain upper energy bound, can be replaced by a theory with a higher upper
bound.
 
Nevertheless, it is possible that the $\chi U\phi$ model in the continuum
limit taken at the tricritical point E defines an interacting theory. The
pursuit of this question requires a better understanding of tricritical points
in four dimensions, as the available experience with such points is restricted
to lower dimensions \cite{LaSa84}. Further obstacles are the necessity to tune
two couplings and the need to extrapolate to the chiral limit, $m_0
\rightarrow 0$. Finally, more insight is needed into strongly coupled and not
asymptotically free nontrivial four-dimensional gauge theories, whose
existence has been recently suggested by numerical investigation of pure U(1)
gauge theory \cite{JeLa96a}.

We remark that the properties of the $\chi U\phi$ model in lower dimensions
are much better accessible. In two dimensions, the numerical evidence strongly
suggests that the continuum limit of the model is equivalent to the
two-dimensional chiral Gross-Neveu model, and is thus renormalizable and
asymptotically free \cite{FrJe96c}. First results in three dimensions
\cite{BaFo98} suggest that the $\chi U\phi$ model belongs to the universality
class of the three-dimensional chiral Gross-Neveu model, which has a
non-Gaussian fixed point. In both cases the continuum limit is obtained on a
whole critical line of chiral phase transitions emerging from the
corresponding Gross-Neveu model obtained in the limit of infinite gauge
coupling, without any use of possible tricritical points. In this sense the
situation in four dimensions is unique, and the experience from lower
dimensions is not applicable.

If the tricritical point E in the four-dimensional $\chi U\phi$ model
defines a renormalizable continuum theory, similar property might be
expected in analogous models with other gauge symmetry groups. For
example, an SU(2) gauge model with scalar and staggered fermion field
in the fundamental representation of SU(2) is known to have at strong
coupling a phase structure very similar to the $\chi U\phi$ model with
the U(1) gauge field \cite{LeShr87a,LeShi86d}.  Therefore we expect
that the model we are studying is generic for a whole class of
strongly coupled gauge models with fermions and scalars in the
fundamental representation.

After describing the $\chi U\phi$ model in the next section, we present our
results as follows: Some preparatory studies of the model in the limit of
infinite bare fermion mass are presented in Section III. In the following
section we demonstrate the existence of the tricritical point. In Section V
the critical and tricritical exponents are estimated by finite size scaling
studies. Spectrum in the continuum limit taken at the point E is discussed in
Section VI. Then we summarize our results and conclude. In the Appendix we give
a detailed definition of the meson propagators and effective Yukawa coupling
we have calculated. 

Preliminary results of this work have been presented in
Refs.~\cite{Fr97a,FrJe97}. An account of our results for the effective Yukawa
coupling between $\pi$ and $F$ is given in a separate paper \cite{FrJe98a}. An
investigation of the $\chi U\phi$ model in the quenched approximation, with
particular emphasis on the role of magnetic monopoles, has been performed in
Ref.~\cite{FrKo98a}. A detailed presentation of the $\chi U\phi$ project in
two, three, and four dimensions can be found in \cite{Fr97b}.  An
investigation of similar models in continuum has been performed by Kondo
\cite{Ko96a}.

%@@@@@@@@@@@@@@@@@@@@@@@@@@@@@@@@@@@@@@@@@@@@@@@@@@@@@@@@@@@@@@@@@@@@@@@
\section{The $\chi U\phi$ model}
%@@@@@@@@@@@@@@@@@@@@@@@@@@@@@@@@@@@@@@@@@@@@@@@@@@@@@@@@@@@@@@@@@@@@@@@
%.......................................................................
\subsection{Action and phase diagram}
%.......................................................................
The four-dimensional lattice $\chi U\phi$ model is defined by the action
\begin{eqnarray}
  S &=& S_\chi + S_U + S_\phi , \\
  S_\chi & = & \frac{1}{2} \sum_x \overline{\chi}_x
    \sum_{\mu=1}^4 \eta_{x\mu} (U_{x,\mu} \chi_{x+\mu} - U^\dagger_{x-\mu,\mu}
    \chi_{x-\mu}) \nonumber \\
    & &+{am_0} \sum_x \overline{\chi}_x \chi_x , \label{eq:schi}\\
  S_U & = & -\beta \sum_P \cos(\Theta_P) , \label{eq:su} \\
  S_\phi & = & - {\kappa} \sum_x \sum_{\mu=1}^4
    (\phi^\dagger_x U_{x,\mu} \phi_{x+\mu} + h.c.). \label{eq:sphi}
\end{eqnarray}
Here $\Theta_P \in [0,2\pi)$ is the plaquette angle, i.\,e.\ the
argument of the product of U(1) link variables $U_{x,\mu}$ along a
plaquette $P$. Taking $\Theta_P = a^2gF_{\mu\nu}$, where $a$ is the
lattice spacing, and $\beta = 1/g^2$, one obtains for weak coupling
$g$ the usual continuum gauge field action $S_U=\frac{1}{4} \int
d^4xF_{\mu\nu}^2$. The staggered fermion field $\chi$ has (real) bare
mass $am_0$ in lattice units and corresponds to four fermion species
in the continuum limit. The scalar field $\phi$ is of fixed modulus,
$|\phi| = 1$.

The model has U(1) global chiral symmetry in the limit $m_0 \rightarrow 0$,
where $m_0$ is the bare fermion mass in physical units, to be defined while
constructing the continuum limit. This is to be distinguished from the limit
$am_0 \rightarrow 0$, allowing explicit chiral symmetry breaking, $m_0 \ne
0$, when $a \rightarrow 0$. Because of this fine difference between $m_0$ and
$am_0$, important in various possible continuum limits, we keep trace of $a$
throughout the paper.

The schematic phase diagram is shown in Fig. \ref{fig:pd4d3}. We recognize
several limit cases of the $\chi U \phi$ model as models interesting
by themselves:
\begin{enumerate} 

\item At $\kappa = 0$ and $am_0 = \infty$, the pure U(1) gauge theory with the
  Wilson action (\ref{eq:su}) and phase transition between the confinement and
  Coulomb phases. Its continuum limit in an extended coupling parameter space
  may be determined by a non-Gaussian fixed point \cite{JeLa96a}.

\item At $\kappa = 0$ and $am_0$ finite, the gauge theory with fermions,
    i.\,e.\ compact QED (\ref{eq:schi}) and (\ref{eq:su}), whose phase
    transition is currently under investigation \cite{CoFr98b}.

\item At $\beta = 0$, i.e. the gauge field being auxiliary, the
  Nambu--Jona-Lasinio (NJL) model, obtained by integrating out the bosonic
  fields \cite{LeShr87a}.  The triviality of this model has been
  recently confirmed in large scale simulations \cite{AlGo95,HaKo97}.

\item At $am_0 = \infty$ and $\kappa$ arbitrary, the compact scalar QED or
  U(1) Higgs model (\ref{eq:su}) and (\ref{eq:sphi}). Its continuum limit at
  strong gauge coupling is Gaussian \cite{AlAz92,AlAz93}.

\end{enumerate}

At strong coupling, $\beta < 1$, the model has three sheets of first order
phase transitions: the two ``wings'' at finite $am_0$, and the sheet at $am_0
= 0$, separating the regions with nonzero chiral condensate of opposite sign.
These three sheets have critical boundary lines E$_{\pm\infty}$E and NE,
respectively. As we shall discuss below, we have verified with solid numerical
accuracy that these 2$^{\rm nd}$ order phase transition lines do indeed
intersect at one point, the tricritical point E. We are not aware of a
convincing theoretical argument why this should be so.

Of most interest is the Nambu phase at $m_0 = 0$, at small $\beta$ and
$\kappa$. Because of confinement, there is no $\phi$-boson, i.\,e.\ charged
scalar, neither fundamental charged $\chi$-fermion in the spectrum. The chiral
symmetry is dynamically broken, which leads to the presence of the neutral
composite physical fermion $F=\phi^\dagger\chi$ with the mass $am_F>0$. It
scales, $am_F\searrow 0$, when the NE line is approached.

Further states include the ``mesons'', i.\,e.\ the fermion-antifermion bound
states: the Goldstone boson $\pi$ with $am_\pi \propto \sqrt{am_0}$, the
scalar $\sigma$, and the vector $\rho$.  ``Bosons'', observed in the
scalar-antiscalar or gauge ball channels are present too. It is in particular
the neutral scalar boson $S$. In the vicinity of the E$_{\pm\infty}$E lines
the same scalar appears both in the $\phi^\dagger-\phi$ and gauge-ball
channels which strongly mix. In the Nambu phase it is natural to interpret $S$
as a gauge ball, as this interpretation holds also deep in the Nambu phase,
when the charged scalar $\phi$ is heavy.

The mass $am_S$ of the $S$-boson vanishes on the lines E$_{\pm\infty}$E,
whereas $am_F$ vanishes on the line NE. Both of them vanish at the tricritical
point E. As their ratio is finite, the continuum limit obtained when
approaching the point E contains both states and is thus different from the
rest of the NE and E$_{\pm\infty}$E lines.

The critical lines E$_{\pm\infty}$E provide another approach to the continuum.
For a nonvanishing $am_0$ the bare mass $m_0$ approaches then infinity and
fermions decouple. The remaining U(1) Higgs model is equivalent to the trivial
$\Phi^4$ theory at the critical endpoint of the Higgs phase transitions
\cite{AlAz92,AlAz93}. This is confirmed by some of our results presented
below.

%.......................................................................
\subsection{Observables and numerical simulations}
%.......................................................................
For the investigation of the tricritical point we use the following
observables:

To localize the Higgs phase transition we use the normalized plaquette and
link energy are defined as
\begin{eqnarray}
  E_{\rm P} &=& \frac{1}{6V}\sum_{x,\,\mu<\nu}\Reop \{U_{x,\mu\nu}\}\;,\\ 
  E_{\rm L} &=& \frac{1}{4V}\sum_{x,\nu}\Reop \{ \phi_x^\dagger
  U_{x,\mu}\phi_{x+\mu} \} \;,
\end{eqnarray}
where V=$L^3T$ is the lattice volume. Following \cite{AlAz93,FrFr95} we
use the perpendicular and parallel components of these energies,
\begin{eqnarray}
  \label{eperp}
  E_\perp &=& E_{\rm L} \cos{\theta} + E_{\rm P} \sin{\theta}\;, \\
  \label{epar}
  E_{||} &=& E_{\rm L} \sin{\theta} - E_{\rm P} \cos{\theta}\;,
\end{eqnarray}
where $\theta$ is the slope of the Higgs phase transition line at the
endpoint in the plane ($\beta, 4/3\kappa$).

For the localization of the chiral phase transition we measure the chiral
condensate
\begin{equation}
  \cbcex= \left\langle\Tr M^{-1} \right\rangle
\end{equation}
via a stochastic estimator, where $M$ is the fermion matrix.

To calculate the mass of the physical fermion we consider the gauge invariant
fermionic field $F_x = \phi^\dagger_x \chi_x$. The mass $am_F$ is measured by
fitting its propagator in momentum space \cite{FrFr95}. The results for the
measurement in configuration space are consistent.

The fermion-antifermion composite states are called ``mesons''. The
corresponding operators and other details are given in \cite{FrJe95a,Fr97b}.
We tried to include also the annihilation part, but failed to obtain
sufficient statistics.

To improve the signal, we also measure the meson propagators with smeared
sources. This required the adaption of the routines, used with Wilson
fermions, to the case of staggered fermions. It is described in Appendix
\ref{app:smmes}. With these smeared sources we have been able to fit the meson
propagators by a one particle contribution at time distances larger than zero.
But the same masses could be obtained if the unsmeared propagators were fitted
with the inclusion of excited states.  The smeared propagators reduce the
errors, however, and in this work we mostly show results obtained by this
method.  Further details of the fitting procedure can be found in
\cite{Fr97b}.

From the propagator of the $\pi$ meson also the pion decay constant $af_\pi$
can be calculated \cite{KiSh87},
\begin{equation}
  \label{fpi}
  af_\pi = \sqrt{Z_\pi}\frac{am_0}{(am_\pi)^2}\;.
\end{equation}
Here $am_\pi$ and $Z_\pi$ are the mass and the wave function renormalization
constant of the $\pi$ meson. We checked, that $af_\pi$ fulfills with excellent
precision the current algebra relation
\begin{equation}
  \label{current}
  (af_\pi am_\pi)^2 = \frac{1}{2} am_0 \cbcex\;.
\end{equation}
This is so even very close to the phase transition, though both $af_\pi$
and $\cbcex$ show rather strong finite size effects there.

For the investigation of the chiral phase transition we also calculate the
susceptibility ratio $R_\pi$, which is defined as the logarithmic derivative of
the chiral condensate \cite{KoKo93a,GoHo94a},
\begin{equation}
  \label{def_rpi}
  R_\pi = \left.\frac{\partial\ln\cbcex}{\partial\ln
 am_0}\right|_{\beta,\kappa} = \left.
 \frac{am_0}{\cbcex}\frac{\partial\cbcex}{\partial am_0}\right|_{\beta,\kappa}\;.
\end{equation}
We measure it as the ratio of zero momentum meson propagators
\begin{equation}
  R_\pi = \frac{C_\sigma(p=0)}{C_\pi(p=0)}
\end{equation}
including the annihilation part of the propagator. This is done by means of a
stochastic estimator and is described in detail in \cite{Fr97b}.

As explained in \cite{KoKo93a}, we expect that close to a critical point the
data could be described by means of the scaling law
\begin{equation}
  \label{scal_rpi}
  R_\pi(t,am_0) = {\cal G}\left(\frac{am_0}{t^\Delta}\right),
\end{equation}
where $t$ is the distance from the critical point (reduced coupling),
$\Delta=\beta+\gamma$ the critical exponent and $\cal G$ a scaling
function. At the critical point,
\begin{equation}
  R_\pi(0,am_0) = {\cal G}(\infty) = \frac{1}{\delta}\,,
\end{equation}
as can be seen by inserting $\cbcex \propto (am_0)^{1/\delta}$ into
equation~(\ref{def_rpi}). At the critical point, $R_\pi$ should be independent
of $am_0$ for sufficiently small $am_0$ (scaling region). In the broken phase,
$R_\pi$ vanishes in the chiral limit, as can been seen easily from the
definition. In the symmetric phase, the $\sigma$ and $\pi$ channels are
degenerate, so that in the chiral limit $R_\pi=1$. For small fixed $t$ a
characteristic behavior is expected, if one varies $am_0$: Because
$\Delta>1$, close to the critical point the curves for $R_\pi$ start for
$am_0=0$ from 0 and 1, respectively, and for increasing $am_0$ approach the
horizontal line $1/\delta$. This will happen the faster the smaller $|t|$ is
(compare fig.~\ref{fig:rpi055}).

Further we consider the scalar and vector bosons, whose operators are defined
as
\begin{eqnarray}
\label{ObosS} 
{\cal O}^{(S)} (t) &=& \frac{1}{L^3} \sum_{\vec{x}} \Reop
  \left\{
    \sum_{i=1}^3 \phi^\dagger_{\vec{x},t} U_{(\vec{x},t), i}
    \phi_{\vec{x}+\vec{\imath},t} \right\} ,\\
  \label{ObosV}
  {\cal O}^{(V)}_{i} (t) &=& \frac{1}{L^3} \sum_{\vec{x}} \Imop
  \left\{ \phi^\dagger_{\vec{x},t} U_{(\vec{x},t), i}
    \phi_{\vec{x}+\vec{\imath},t} \right\}, \\
  &  & i = 1,2,3. \nonumber
\end{eqnarray}
The masses $am_S$ and $am_V$ of the scalar and vector bosons are calculated
from the corresponding correlation functions in configuration
space\footnote{To reduce the statistical fluctuations in the determination of
  $am_S$, calculating the propagator we subtract the momentum zero propagator
  before the determination of the error (average over the propagator).}.

In the same way we also measure the gauge invariant combinations of the
gauge fields, which we call gauge balls, in analogy to QCD glue balls. We
define two operators with the quantum numbers $0^{++}$ and $1^{+-}$:
\begin{eqnarray}
\label{O0++}
  {\cal O}^{(0^{++})} (t) &=& \frac{1}{L^3} \sum_{\vec{x}} \Reop \left\{
    \sum_{i=2}^3\sum_{j=1}^{i-1} U_{(\vec{x},t),ij}
  \right\},\\
\label{O1+-}
  {\cal O}^{(1^{+-})}_{i} (t) &=& \frac{1}{L^3} \sum_{\vec{x}} \Imop \left\{
    U_{(\vec{x},t),jk} \big|_{i\neq j\neq k\neq i}
  \right\}, \\
  &  & i  = 1,2,3. \nonumber
\end{eqnarray}
The masses $am_{\text{G}}$ of the gauge-balls are calculated in analogy to the
boson masses by means of the propagators in the configuration space.

We have observed mixing of the $S$ boson and the $0^{++}$ gauge ball by means
of the two-point function
\begin{equation}
  G^{(S,G)}(t)=\left\langle {\cal O}^{(S)}(0){\cal O}^{(0^{++})}(t)
  \right\rangle .
\end{equation}

We have also measured the effective Yukawa coupling $y_{\text{R}}$ between the
neutral fermion $F$ and the $\pi$ meson. This is done in analogy to
\cite{GoHo95}, and the used operators are described in appendix \ref{app:yr}.
A detailed discussion of our results is given in \cite{FrJe98a} and summarized
in the conclusion of the present paper.

%@@@@@@@@@@@@@@@@@@@@@@@@@@@@@@@@@@@@@@@@@@@@@@@@@@@@@@@@@@@@@@@@@@@@@@@
\section{Limit of infinite bare fermion mass}
%@@@@@@@@@@@@@@@@@@@@@@@@@@@@@@@@@@@@@@@@@@@@@@@@@@@@@@@@@@@@@@@@@@@@@@@
\label{sec:higgs}

%.......................................................................
\subsection{U(1) Higgs model and chiral phase transition in the quenched
  approximation} 
%.......................................................................
For $am_0 = \infty$, the \chup-model reduces to the U(1) Higgs model with
$|\phi| = 1$ on the lattice, (\ref{eq:su}) and (\ref{eq:sphi}). This model has
been investigated in the eighties (for a review see e.g. \cite{Je86}) and with
modern methods in \cite{AlAz92,AlAz93}. Its phase diagram is represented by
the front face of Fig. \ref{fig:pd4d3}. It has the Coulomb phase at small
$\kappa$ and large $\beta$, the rest being the confinement-Higgs phase. The
line of Higgs phase transitions E$_\infty$S$_\infty$ is first order except the
points E$_\infty$ and S$_\infty$. The continuum limit at the critical endpoint
E$_\infty$ corresponds most probably to a trivial scalar field theory
\cite{AlAz92,AlAz93}.

When dynamical fermions with $am_0>0$ are included, the phase diagram remains
roughly the same, except that the confinement-Coulomb phase transition and the
endpoint E$_{am_0}$ shift to smaller $\beta$. The endpoints then form the
critical line E$_{\infty}$E. It is natural to expect that this line, except
the tricritical point E, remains in the same universality class as the point
E$_\infty$. Our results confirm this expectation.

When quenched fermions with small $am_0$ are included into the Higgs
model, a line of chiral phase transitions appears in the otherwise
unchanged phase diagram of the Higgs model. It was realized already in
the first investigations of the Higgs model with fermion, that full
and quenched models have a very similar phase diagram
\cite{LeShr87a,LeShi86d}. This includes the observation that the
chiral phase transition line runs within numerical accuracy into the
critical endpoint of the Higgs phase transition line. The phase
diagram of the quenched model looks thus similar to the $am_0 = 0$
plane of Fig.  \ref{fig:pd4d3}.

This similarity suggests that it might be instructive to study the \chup-model
in the quenched approximation. In Ref. \cite{FrKo98a} a quenched investigation
of the interplay of chiral phase transition and the monopole percolation was
performed. It seems that there might be an interplay of both transitions at an
intermediate $\beta$ on the N$_\infty$E$_\infty$ line, possibly with
nontrivial exponents. Around the points N$_\infty$ and E$_\infty$ the chiral
and the percolation transitions appear to be separated, however.

%.......................................................................
\subsection{Scaling behavior at the endpoint E$_\infty$}
%.......................................................................
\label{sec:higgsend}
We begin by investigating the endpoint E$_\infty$. We want to gain
experience and check the reliability of the determination of critical
exponents by means of Fisher zeros. We later apply this method at
finite $am_0$ for the scaling investigation along the E$_\infty$E line
and compare the results with those at E$_\infty$.

\begin{figure}[tbp]
  \begin{center}
    \psfig{file=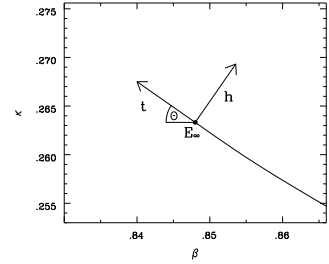,height=6cm,angle=90}\hspace*{-10mm}%
    \caption{%
      Definition of the reduced couplings $t$ and $h$ and of the angle
      $\theta$. The point E$_\infty$ is the endpoint of the Higgs phase
      transition line. The scale of the $\kappa$ axis is 4/3 times larger than
      that of the $\beta$ axis.}
    \label{fig:qpar}
  \end{center}
\end{figure}

The scaling behavior at the endpoint of the Higgs phase transition line was
determined in \cite{AlAz92,AlAz93} along the first order Higgs transition
line. It was found that the endpoint is described by mean field exponents. We
investigate the scaling behavior approaching $E_\infty$ in different
directions. For this purpose it is useful to introduce the following reduced
couplings (Fig.~\ref{fig:qpar}):\\ 
%%
%FFFFFFFFFFFFFFFFFFF Fig. 2 FFFFFFFFFFFFFFFFFFFFFFFFFFFFFFFFFFFFFFFFFF
\hspace*{7mm}\begin{tabular}[t]{@{}ccl}
  $t$ &:& parallel to the phase transition line of 1$^{\rm st}$ order\\
  $h$ &:& perpendicular to the phase transition line.
\end{tabular}\\
Here perpendicular is understood in the same sense as in eqs.\eqref{eperp} and
\eqref{epar}, so that
\begin{eqnarray}
  \label{def_t}
  t &=& - (\beta-\beta_c) \cos\theta +
  {\textstyle\frac{4}{3}}(\kappa-\kappa_c)\sin\theta\, , \\ 
  \label{def_h}
  h &=& \phantom- (\beta-\beta_c) \sin\theta\, +
  {\textstyle\frac{4}{3}}(\kappa-\kappa_c)\cos\theta\, ,
\end{eqnarray}
and therefore
\begin{equation}
  S = -6V(t E_{||} + h E_\perp) + S_{E_\infty}\;.
\end{equation}
The letters $t$ and $h$ have been chosen in analogy to temperature and
external field in magnetic systems.

We introduce critical exponents $\nu$ and $\tilde\nu$ of the correlation
length for both directions,
\begin{equation} \label{NU}
  \xi \propto |t|^{-\nu}|_{h=0}\, ,
\end{equation}
and
\begin{equation} \label{NU_TILDE}
  \xi \propto |h|^{-\tilde{\nu}}|_{t=0}\, ,
\end{equation}
$\xi$ being the correlation length diverging at E$_\infty$.

To understand the relation between $\nu$ and $\tilde\nu$ we assume the
equation of state
\begin{equation}
  \label{scal_higgse}
  \xi = |t|^{-\nu} F\left(\frac{|h|}{|t|^\Delta}\right)\;,
\end{equation}
with the scaling function $F$ and $\Delta=\beta+\gamma$. Introducing
$\tilde{F}(x)= x^{\nu} F(x^\Delta)$ it can be rewritten as
\begin{equation}
  \label{scal_higgset}
  \xi = |h|^{-\nu/\Delta}
  \tilde{F}\left(\frac{|h|^{1/\Delta}}{|t|}\right)\;.
\end{equation}
Assuming $\tilde{F}(\infty)<\infty$ this means
$\tilde\nu=\nu/\Delta$.

The scaling behavior (\ref{NU_TILDE}) is expected in the general
direction $t=c \cdot h$, because $\Delta>1$, and therefore
\begin{equation}
  \frac{|h|^{1/\Delta}}{|t|} = h^{1/\Delta-1}/c \rightarrow\infty
\end{equation}
for $h\rightarrow 0$ and, accordingly, $t\rightarrow 0$. This makes clear that
it is not important to choose $h$ perpendicular to $t$. Only the $t$-direction
($h=0$) is special as it is tangential to the phase transition line and thus
described by the scaling law (\ref{NU}).

The mean field values of the exponents $\beta=1/2$, $\gamma=1$, and $\nu=1/2$
correspond to $\tilde{\nu}=1/3$.

To determine the critical exponent of the correlation length we measure the
scaling behavior of the edge singularity in the complex coupling plane (Fisher
zero)\cite{Fi64}. From scaling arguments for the free energy we expect for
the first zero $z_1$:
\begin{equation}
  \label{scal_ntriv}
  \left.\Imop z_1(L)\right|_{t=0} = A \cdot L^{-1/\tilde{\nu}}\;.
\end{equation}
As all directions which are not tangential to the phase transition line are
equivalent, we expect the same exponent $\tilde\nu$ also if we fix
$\beta=\beta_{\rm E_\infty}$ or $\kappa=\kappa_{\rm E_\infty}$. This was
verified in \cite{Fr97a}. Fixing one of the couplings is particularly
convenient for the necessary analytic extra\-polations into the complex plane.
That is done by means of the multihistogram reweighting method \cite{FeSw89}.

\begin{figure}[tbp]
  \begin{center}
    \parbox{8mm}{(a)}%
    \parbox[c]{.9\hsize}{\fdfig{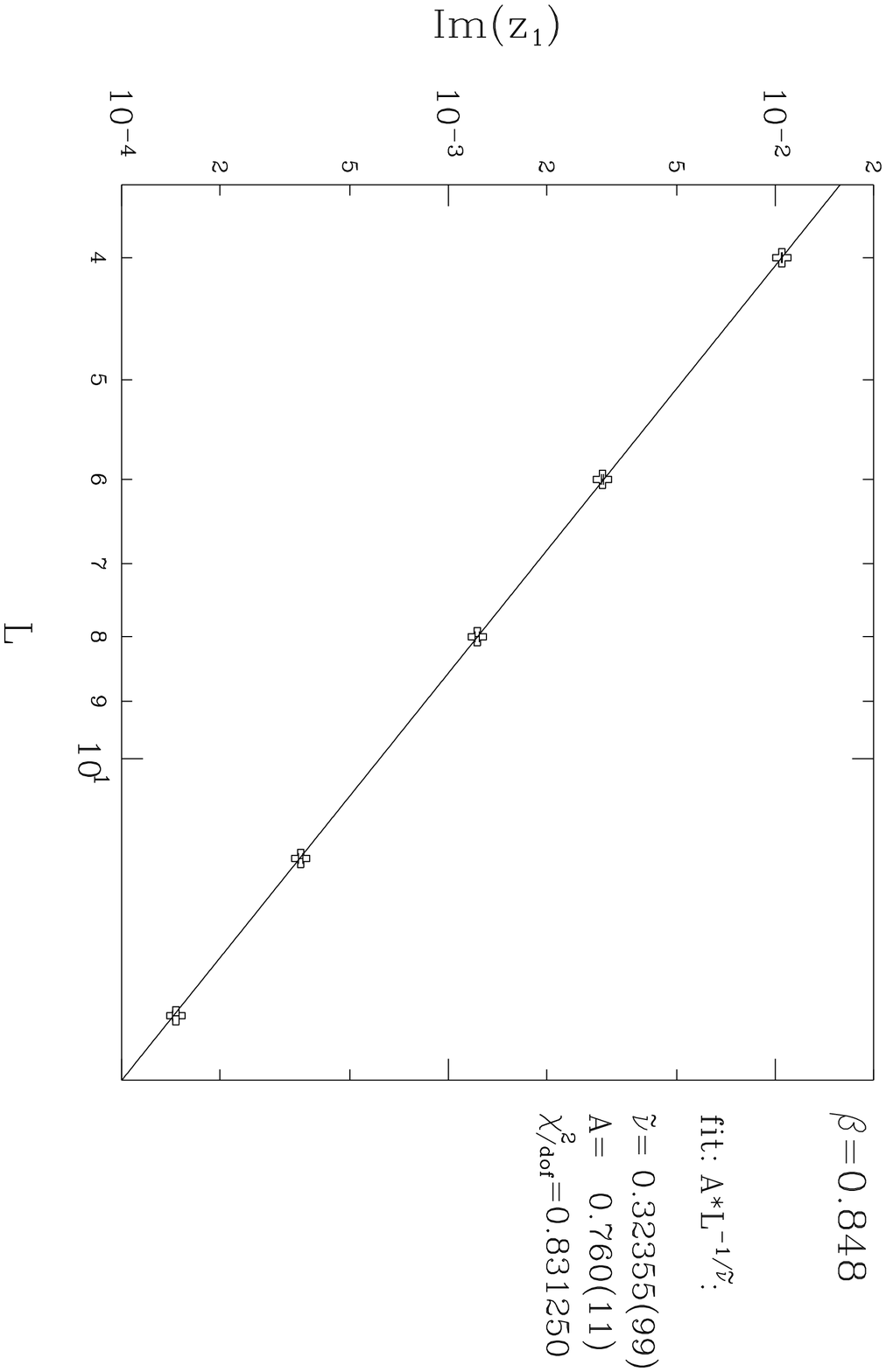}}\\[3mm]
    \parbox{8mm}{(b)}%
    \parbox[c]{.9\hsize}{\fdfig{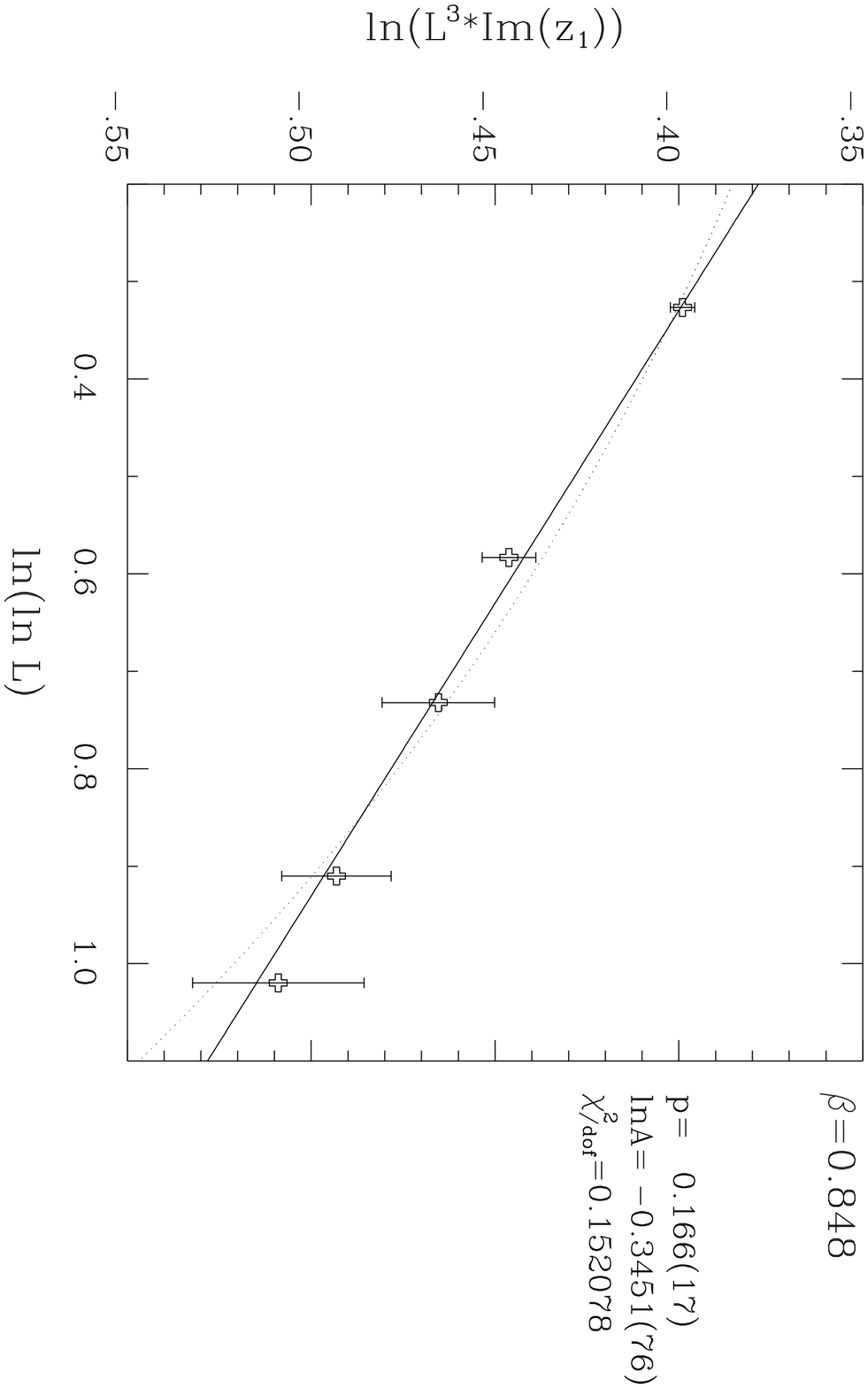}}
    \caption{%
      (a) Scaling behavior of the edge singularity for $\beta=0.848\simeq
      \beta_{\text{E}_\infty}$ in the U(1) Higgs model on $L^4$ lattices. The
      small deviations from the trivial scaling behavior with $\tilde\nu=1/3$
      are resolved in (b). Here $\ln(L^3\Imop z_1)$ is shown as a function of
      $\ln(\ln L)$. The fit from Fig.~(a) is shown dotted. The value for
      $\beta_{\text{E}_\infty}$ has been taken from \protect\cite{AlAz93}.}
    \label{fig:qfisher1}
  \end{center}
\end{figure}

We present here the scaling investigation we did for
$\beta=0.848\approx\beta_{\rm E_\infty}$.  Fig.~\ref{fig:qfisher1}a shows a
nice scaling behavior for all lattice sizes with the critical exponent
$\tilde\nu=0.3236(10)$. This value is very close the mean field exponent
$\tilde\nu=1/3$.

%%
%FFFFFFFFFFFFFFFFFFF Fig. 3 FFFFFFFFFFFFFFFFFFFFFFFFFFFFFFFFFFFFFFFFFF

The small deviations outside the error bars are probably due to
logarithmic corrections, as they are expected at Gaussian fixed
points.  To verify this we follow the idea from \cite{KeLa93} and factor
out the leading power law $L^{-1/\tilde{\nu}}$. If the scaling
behavior has the form
\begin{equation}
  \label{scal_triv}
  L^3 \left.\Imop z_1(L)\right|_{t=0} = A \cdot (\ln L)^{-p}\;,
\end{equation}
we expect in the $\ln\ln$ plot a strait line with the slope $-p$. The data
shown in Fig.~\ref{fig:qfisher1}b are very well described by a straight line
with $p \simeq 0.17$. The $\chi^2$ is smaller than that with a fit by means of
the equation (\ref{scal_ntriv}). However, we have not investigated how far
these results, and especially $p$ depend on the precise knowledge of the
critical point.

A similar value for the exponent $\tilde\nu$ was also measured in
the SU(2) Higgs model \cite{BoEv90b,Bo90}, but then it was not
realized that this value is actually compatible with a Gaussian fixed
point. Now we can conclude that the U(1) and SU(2) Higgs models have a very
similar scaling behavior at the endpoint.

We have also determined the critical exponent $\tilde\nu$ by means of the
finite size scaling of the specific heat and of some cumulants \cite{Fr97b}.
These less precise methods confirm the results presented here. Also the shift
exponent $\lambda$ has turned out to be compatible with $1/\tilde\nu$ for the
investigated observables \cite{Fr97a,Fr97b}.

%@@@@@@@@@@@@@@@@@@@@@@@@@@@@@@@@@@@@@@@@@@@@@@@@@@@@@@@@@@@@@@@@@@@@@@@
\section{Existence and position of the tricritical point $E$}
%@@@@@@@@@@@@@@@@@@@@@@@@@@@@@@@@@@@@@@@@@@@@@@@@@@@@@@@@@@@@@@@@@@@@@@@

\subsection{General properties of tricritical points}

To present our investigations of the tricritical point E, we first summarize
the relevant general properties of tricritical points and define the
exponents.  In notation we follow Griffiths \cite{Gr73}.

\begin{figure}
  \begin{center}
    \psfig{file=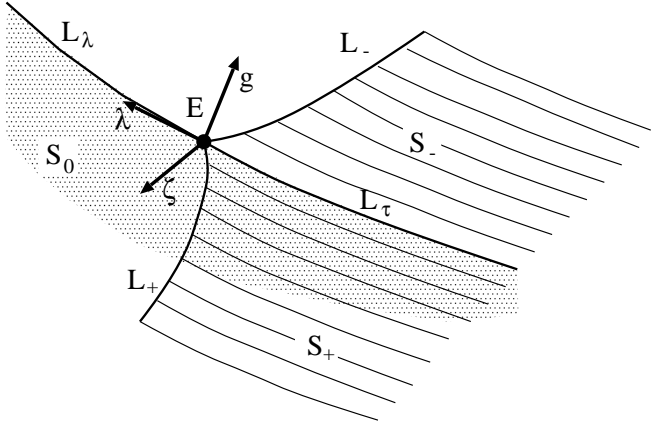,width=\hsize}
    \vspace{3mm}
    \caption[xxx]{%
      Schematic phase diagram of the first order PT planes and the critical
      lines (L$_\pm$, L$_\lambda$) in the vicinity of the tricritical point E.
      L$_\tau$ is the line of triple points. The L lines correspond the lines
      in Fig.~\ref{fig:pd4d3} as follows: L$_\lambda\,\hat=\,\mbox{NE}$,
      L$_\pm\,\hat=\,\mbox{E$_{\pm\infty}$E}$, L$_\tau\,\hat=\,\mbox{ET}$. A
      local coordinate system is shown.}
    \label{fig:param_tri}
  \end{center}
\end{figure}

In the vicinity of a tricritical point it is usual to choose the following
orthogonal coordinate system (Fig.~\ref{fig:param_tri}):\\ 
\begin{tabular}[t]{@{}ccp{80.5mm}@{}}
  $\lambda$ &:& tangential to the first order PT line in the symmetry plane,\\
  $g$ &:& perpendicular to the PT line inside the symmetry plane,\\
  $\zeta$ &:& perpendicular to the symmetry plane.
\end{tabular}\\
In the symmetry plane $m_0=0$, these definitions are analogous to those in the
Higgs model (sec. \ref{sec:higgsend}), $\lambda$ and $g$ corresponding to $t$
and $h$, respectively.

In the phase diagram there are four special lines, which we denote following
\cite{LaSa84}: the chiral PT line NE (second order) in the symmetry plane
($\lambda>0$) is lambda line L$_\lambda$, its continuation in the symmetry
plane, on which three first order phase transition sheets meet ($\lambda<0$),
is triple line L$_\tau$, and the two lines of endpoints outside the symmetry
plane are wing critical lines L$_+$ and L$_-$. The first order PT plane below
the lines L$_\lambda$ and L$_\tau$ in the symmetry plane is denoted S$_0$, and
the two wings of Higgs phase transitions are S$_+$ and S$_-$. Because of the
$\pm am_0$ symmetry we use in the following only the index $+$.
%%
%FFFFFFFFFFFFFFFFFFF Fig. 4 FFFFFFFFFFFFFFFFFFFFFFFFFFFFFFFFFFFFFFFFFF

\begin{table}
  \begin{center}
    \renewcommand{\arraystretch}{1.4}
    \begin{tabular}{c|c|l|c}
      exp. & \cite{LaSa84} & definition & class.\\
           &               &            & value \\\hline $\alpha_t$ &
      $\alpha_t$ & $\frac{\partial E_\perp}{\partial g} \propto
      g^{-\alpha_t}$, $\lambda=\zeta=0$ & $\frac{1}{2}$ \\
      $\alpha_u$ & $\alpha$ & $\frac{\partial E_{||}}{\partial \lambda}
      \propto \lambda^{-\alpha_u}$, $g=\zeta=0$ & -1 \\
      $\beta_t$ & $\beta_t$ & $\cbcex \propto |g|^{\beta_t}$,
      $\lambda=\zeta=0$
      & $\frac{1}{4}$\\
      $\beta_u$ & $\beta_2$ & $\Delta E_\perp \propto |\lambda|^{\beta_u}$,
      $g=\zeta=0$,
      $\lambda<0$ & 1\\
      $\delta_t$ & $\delta$ & $\cbcex \propto \zeta^{1/\delta_t}$,
      $\lambda=g=0$ & 5\\
      $\delta_u$ & $1/\beta_{2t}$ & $|E_\perp-E_{\perp c}| \propto
      |g|^{1/\delta_u}$,
      $\lambda=\zeta=0$ & 2\\
      $\nu_u$ & & $\xi_+ \propto |\lambda|^{-\nu_u}$, $g=\zeta=0$ & 1\\
      $\tilde\nu_u$ & & $\xi_+ \propto |g|^{-\tilde\nu_u}$, $\lambda=\zeta=0$
      & $\frac{1}{2}$\\
      $\nu_t$ & & $\xi_\lambda \propto |g|^{-\nu_t}$, $\lambda=\zeta=0$ & $\frac{1}{2}$\\
      $\phi$ & $\phi$ & $g \propto |\lambda|^\phi$ at the lines L$_\lambda$ and
      L$_\tau$, $\zeta=0$ & 2\\
      $\phi\Delta_t$ & $\Delta$ & $\zeta \propto |\lambda|^{\phi\Delta_t}$ at
      the lines L$_+$ & $\frac{5}{2}$\\
      $\Delta_u$ & & $\Delta_u=\beta_u\delta_u$ & 2\\
    \end{tabular}
    \caption{Exponents at the tricritical point. In the second column the
      notation of \protect\cite{LaSa84} is given. The classical value is derived in
      three dimensions. Its applicability to four dimensions is discussed in
      the text.}
    \label{tab:exp}
  \end{center}
\end{table}

The most important exponents and the defining scaling behavior are summarized
in table \ref{tab:exp}.  Comparing a metamagnet to our model, the staggered
magnetization corresponds to the chiral condensate and the magnetization to
the energy $E_\perp$. The unfortunate fact is that the symmetry plane, which
is of major use in the study of metamagnets, corresponds to the chiral limit
$m_0 = 0$, which is difficult to approach in numerical simulations with
fermions coupled to a gauge field.

The two sets of exponents with index $t$ and $u$ are defined in analogy to the
exponents on the adjacent critical lines: The set with the index $t$
(tricritical exponents) is defined in analogy to the exponents along the
chiral PT line NE (L$_\lambda$ line). The set with the index $u$ (subsidiary
exponents) is defined in analogy to the exponents along the line of Higgs
endpoints E$_\infty$E (L$_+$ line). In general, the exponents at the
tricritical point are different from those at the adjacent lines. We denote
the diverging correlation lengths and the exponents on the lines L$_\lambda$
and L$_+$ by the indices $\lambda$ and $+$, respectively.

The tricritical point can be considered as a special point of both the
L$_\lambda$ and the L$_+$ critical lines. Both the correlation length
$\xi_\lambda$ diverging at the L$_\lambda$ line and the correlation length
$\xi_+$ diverging at the L$_+$ line are critical there. Nevertheless, in
general, $\xi_\lambda$ and $\xi_+$ are to be distinguished also at the
tricritical point. In our case $\xi_\lambda = 1/am_F$ and $\xi_+ = 1/am_S$.

Our results strongly indicate (sec.~\ref{sec:scal_sf}), that at the point E,
$\xi_\lambda\propto\xi_+$ on all paths into the point E. This proportionality
seems to hold also for all other observed correlation lengths (inverse masses)
which diverge at the point E. Thus there seems to be only one scaling law and
$\tilde\nu_u=\nu_t$. This is a generic property of tricritical points. It
makes possible to use at the tricritical point the finite size scaling theory
quite in analogy to the adjacent critical lines.

In three dimensions it is usual that tricritical points have a large region of
dominance. In analogy, near the tricritical point we expect to find at some
distance from the second order PT lines already the scaling behavior described
by the tricritical exponents. Such a crossover phenomenon was investigated for
example in \cite{La76}. A similar effect might be expected also on small
lattices in the immediate vicinity to the PT lines. It is not excluded,
however, that in four dimensions the tricritical points are much less dominant
than in three-dimensional models. To the best of our knowledge, tricritical
points have not yet been investigated numerically in four dimensions.

For the singular part of the free energy $F$ usually the following scaling
relations are assumed \cite{LaSa84}:
\begin{eqnarray}
  F_{\rm sing}(\lambda,g,\zeta) &=& |\lambda|^{2-\alpha_u}
  {\cal F}^{(\pm)}\left(\frac{g}{|\lambda|^{\phi}},
    \frac{\zeta}{|\lambda|^{\phi\Delta_t}}\right),\\
  F_{\rm sing}(\lambda,g,\zeta) &=& |g|^{2-\alpha_t}
  {\cal F}_1^{(\pm)}\left(\frac{\lambda}{|g|^{1/\phi}},
    \frac{\zeta}{|g|^{\Delta_t}}\right)\;.
\end{eqnarray}
For such systems several relations between the exponents can be derived
\cite{Gr73}:
\begin{eqnarray}
  \alpha_t &=& 1-1/\delta_u \, ,\\
  \Delta_u &=& \phi\, ,\\
  2-\alpha_u &=& \phi(2-\alpha_t)\, ,\\
  \nu_u &=& \phi\nu_t\\
  1+\delta_t &=& (2-\alpha_t)/\beta_t\, ,\\
  \label{scal_Deltat}
  \Delta_t &=& \beta_t\delta_t = (2-\alpha_u)/\phi\cdot\delta_t/(1+\delta_t)\, ,\\
  \label{scal_betat}
  \beta_t &=& (2-\alpha_t)/(1+\delta_t)\, .
\end{eqnarray}
In our work we use in particular the last two of these relations.

Only four exponents are independent. With the assumption that the hyperscaling
relation
\begin{equation}
  \label{scalt_joseph}
  \alpha_t = 2 - d \nu_t
\end{equation}
holds, only three independent exponents remain.

Unfortunately, theory of tricritical points in four dimensions is developed
only insufficiently. The Ginzburg criterion indicates that for all dimensions
$d \geq 3$, tricritical point can be described by the classical exponents.
This would correspond in four dimensions to a violation of the hyperscaling
relation \cite{LaSa84}.

%.......................................................................
\subsection{Two diverging correlation lengths}
%.......................................................................
\label{sec:extri}
For the existence of the tricritical point E in the \chup\ model the chiral
and Higgs phase transition must meet at one point in the $m_0 = 0$ plane.
Since there is no theoretical understanding for the interplay between both
transition, the existence of such a point has to be demonstrated numerically.

\begin{figure}
  \begin{center}
    \fdfig{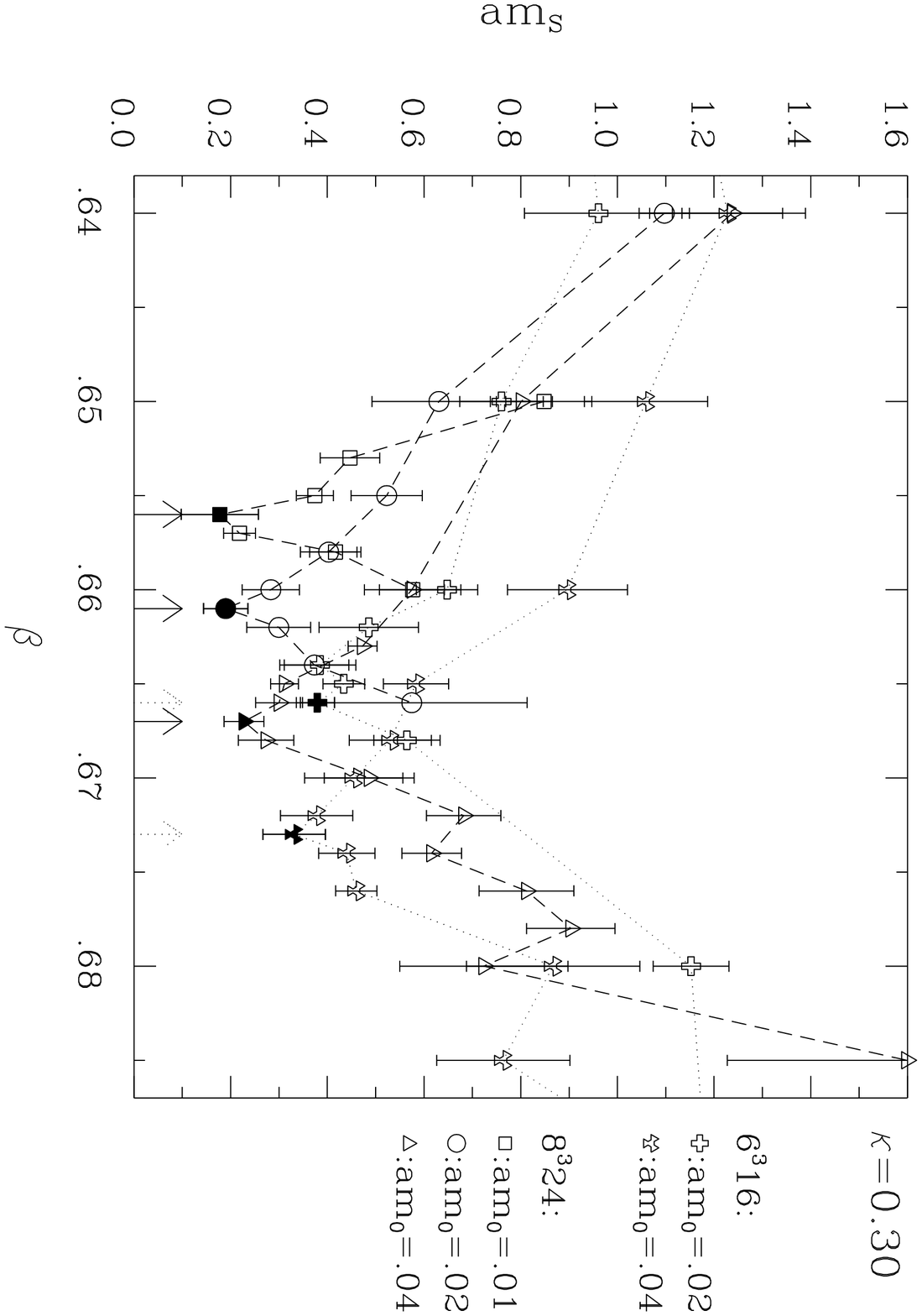}
    \caption{%
      Boson mass $am_S$ as function of $\beta$ for different $am_0$ and
      lattice sizes at $\kappa=0.30\approx \kappa_{\rm E}$. The arrows
      indicate the minima of $am_S$.}
    \label{fig:bosc1}
    \vspace{8mm}
%FFFFFFFFFFFFFFFFFFF Fig. 6 FFFFFFFFFFFFFFFFFFFFFFFFFFFFFFFFFFFFFFFFFF
    \fdfig{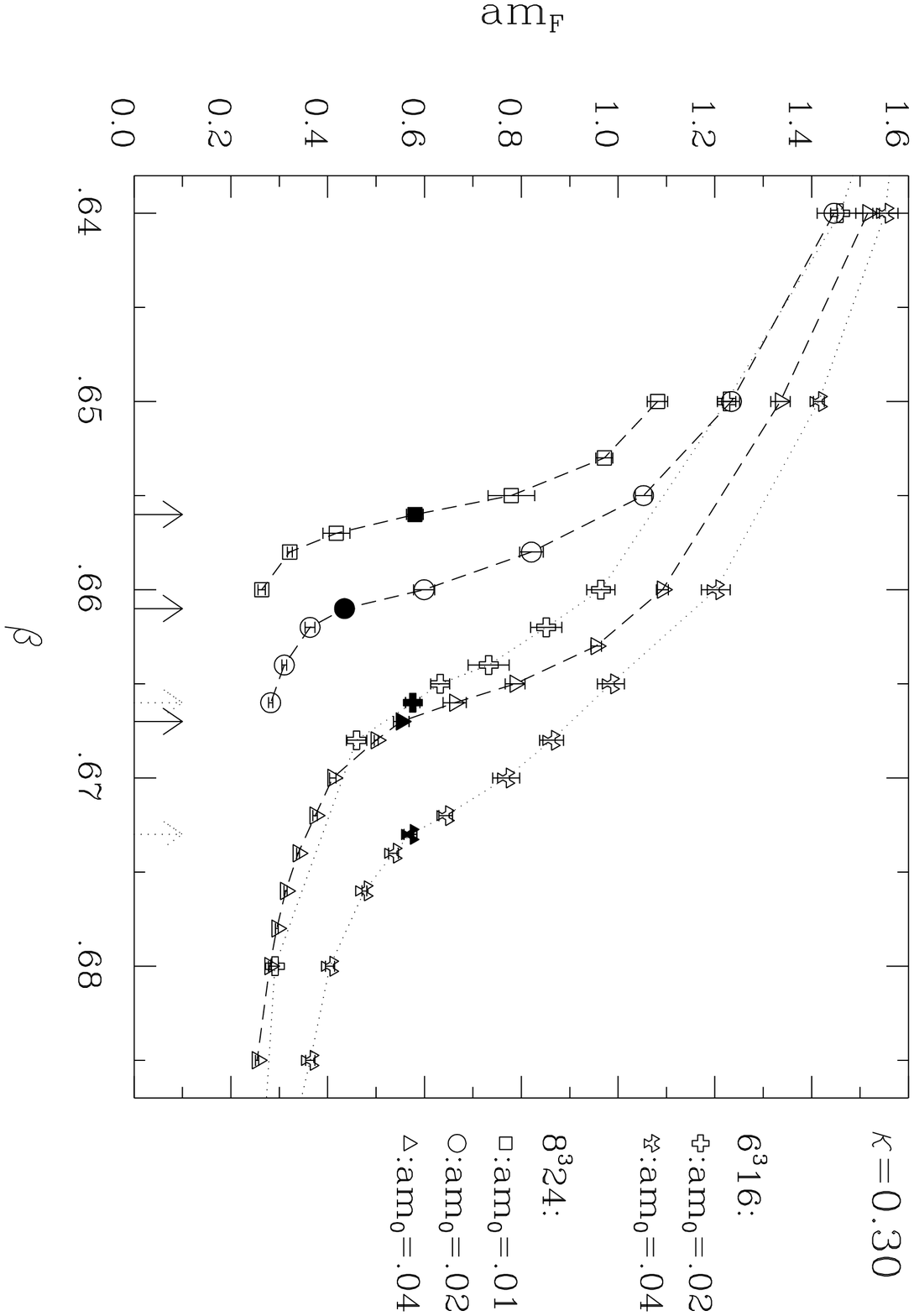}
    \caption[xxx]{%
      Fermion mass $am_F$ as function of $\beta$ for different $am_0$ and
      lattice sizes at $\kappa=0.30$. The minima of the boson mass $am_S$ are
      marked with full symbols and arrows (cf.\ Fig.~\ref{fig:bosc1}).}
    \label{fig:fimp1}
  \end{center}
\end{figure}

To give a first impression, Fig.~\ref{fig:bosc1} shows the mass of the scalar
boson $am_S$ in the vicinity of the tricritical point for $am_0=0.01$, 0.02,
and 0.04. This mass has been obtained from the $\phi^\dagger-\phi$ correlation
function (\ref{ObosS}). It has a pronounced minimum (arrows in
Fig.~\ref{fig:bosc1}) for each $am_0$. Its minimal value on the $6^316$
lattice is about 0.35, and 0.2 on the $8^324$ lattice. The significant
decrease with increasing lattice size indicates that the mass in the infinite
volume vanishes, and the correlation length $\xi_+ = 1/am_S$ thus diverges.
The vanishing of $am_S$ at some $\beta$, $\kappa$ for any finite $am_0$
corresponds to the E$_\infty$E line of endpoints E$_{am_0}$ of the Higgs phase
transitions. 

%FFFFFFFFFFFFFFFFFFF Fig. 5 FFFFFFFFFFFFFFFFFFFFFFFFFFFFFFFFFFFFFFFFFF
The fermion mass $am_F$ descends steeply (Fig.~\ref{fig:fimp1}) at the
position of the minimum of the boson mass for the same $am_0$ and volume. In
the broken phase, the curves for different volumes slowly approach each other.
In the symmetric phase, the values of $am_F$ achieve small finite values which
should vanish in the chiral limit and infinite volume.

Figures \ref{fig:bosc1} and \ref{fig:fimp1} show that changes of
volume and $am_0$ result in a shift of the minimum in $\beta$. The same holds for
$\kappa$. So there is little hope to extrapolate the data at fixed $\beta$ and
$\kappa$ into infinite volume and chiral limit. As usual for tricritical
points, a fine tuning of both couplings is required. In this work
we assume that the limited precision of the position of the tricritical point
we have achieved is outweighted by a sufficiently large domain of dominance of
this point.

%.......................................................................
\subsection{Position of the tricritical point E}
%.......................................................................
\label{sec:latheat}
To localize the point E we determine the positions of the endpoints E$_{am_0}$
for small $am_0$ and extrapolate them to $am_0=0$. It is difficult to control
the uncertainties in every step of this procedure.

\begin{figure}
  \begin{center}
    \fdfig{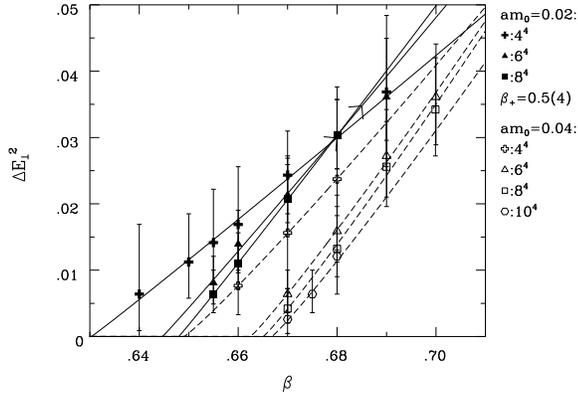}
    \caption[]{%
      Square of the latent heat of the energy $E_\perp$ for $am_0=0.02$ and
      $am_0=0.04$ for lattices of the size $4^4$ to $8^4$ and $10^4$, resp..
      The mean-field exponent $\beta_+=\frac{1}{2}$ corresponds to a strait
      line. We have described the data for each $am_0$ with one common
      $\beta_+$ in eq.~\eqref{scal_de} for the different lattice sizes.}
    \label{fig:latheat}
  \end{center}
\end{figure}

At nonzero $am_0$ we proceed in analogy to \cite{AlAz93}. We determine the
latent heat of $E_\perp$ in the S$_+$ plane.  Fig.~\ref{fig:latheat} shows the
latent heat $\Delta E_\perp$ as a function of $\beta$ for two $am_0$ and
different lattice sizes. The values have been obtained by reweighting the data
to the coupling where both maxima have the same height. Then the distance of
these maxima in the histogram and the uncertainty have been estimated, as the
data is not sufficient for a fitting procedure.
%%
%FFFFFFFFFFFFFFFFFFF Fig. 7 FFFFFFFFFFFFFFFFFFFFFFFFFFFFFFFFFFFFFFFFFF

We expect a scaling of $\Delta E_\perp$ for fixed $am_0$ and fixed lattice
size of the form
\begin{equation}
  \label{scal_de}
  \Delta E_\perp \propto t^{\beta_+} \propto
  \left(\beta-\beta_{pc}(L,am_0)\right)^{\beta_+}\;.
\end{equation}
The index $+$ denotes the magnetic exponent on the E$_\infty$E line.
In this procedure it is assumed that the dominant contribution due to the
finite volume can be absorbed in a volume dependent pseudocritical coupling
$\beta_{pc}(L,am_0)$. A better method would have been to extrapolate the
latent heat first into the infinite volume and to investigate scaling
afterwards. But for such an analysis the quality of our data is not
sufficient.

In the quadratic plot in Fig.~\ref{fig:latheat} we expect a straight line for
the classical value $\beta_+=\frac{1}{2}$, which was observed for
$am_0=\infty$ \cite{AlAz93}. Our data at small $am_0$ are well compatible with
this expectation.

\begin{figure}
  \begin{center}
    \fdfig{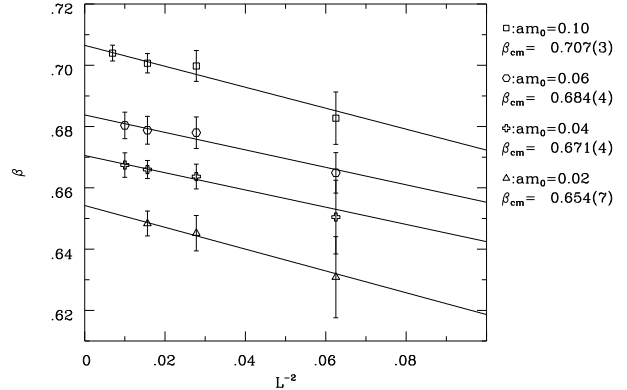}
    \caption{%
      Pseudocritical couplings $\beta_{pc}(L)$ as function of $L^{-2}$ for
      different $am_0 \leq 0.10$. The fit is an extrapolation into infinite
      volume assuming the exponent value $\nu_+=0.5$.}
    \label{fig:scalbpc}
  \end{center}
\end{figure}

A fit with free $\beta_+$ gives a value of $\beta_+=0.5(3)$ for $am_0=0.04$,
and $\beta_+=0.5(4)$ for $am_0=0.02$. The probably overestimated error for
$\Delta E_\perp$ results in an overestimated error of $\beta_+$.  In fact our
data do not have the necessary quality to investigate the scaling
(\ref{scal_de}) with a free exponent $\beta_+$. Therefore, we fit them with
fixed $\beta_+=0.5$. For $\beta_{pc}(L,am_0)$ obtained in this way we assume
scaling as in the Higgs model \cite{AlAz93} with $\nu_+=0.5$:
\begin{equation}
  \label{BETA_PC}
  \beta_{pc}(L,am_0) - \beta_c(am_0) \propto L^{-1/\nu_+}\;.
\end{equation}
Our data are compatible with this assumption (Fig.~\ref{fig:scalbpc}).
%%
%FFFFFFFFFFFFFFFFFFF Fig. 8 FFFFFFFFFFFFFFFFFFFFFFFFFFFFFFFFFFFFFFFFFF

\begin{figure}
  \begin{center}
    \fdfig{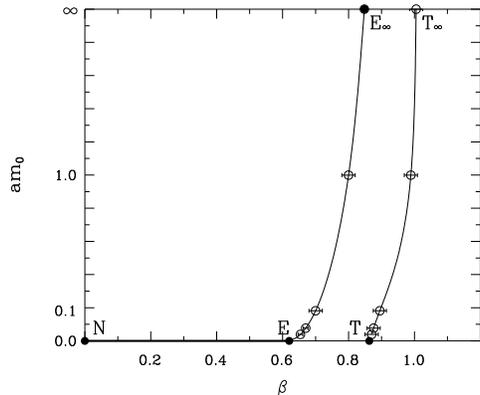}
    \caption[xxx]{%
      Projection of the critical line of endpoints E$_\infty$E and the line of
      triple-points TT$_\infty$ onto the $\beta$-$am_0$-plane. (In
      Fig.~\protect\ref{fig:pd4d3} this corresponds to a view from below for
      $am_0 \ge 0$.) Shown is also the critical line NE of chiral
      phase transitions. The points have been determined on $6^4$ and $8^4$
      lattices. The error bars reflect the uncertainty of the extrapolation
      into the infinite volume. }
    \label{fig:pd4proj}
  \end{center}
\end{figure}

The so determined points $\beta_c(am_0)$ of the E$_\infty$E line are listed in
table~\ref{tab:lineee} and plotted in Fig.~\ref{fig:pd4proj}. The point at
$am_0=1.00$ was only estimated and the uncertainty included in error bars \cite{Fr97b}.
\begin{table}
  \begin{center}
    \begin{tabular}{ccc}
      $am_0$ & $\beta_c$ & $\kappa_c$ \\ \hline
      0.02 & 0.654(7) & 0.304(5) \\
      0.04 & 0.671(4) & 0.296(4) \\
      0.06 & 0.684(4) & 0.292(3) \\
      0.10 & 0.707(3) & 0.285(3) \\
      1.00 & 0.80(2) & 0.28(1) \\
      $\infty$ & 0.848(4) & 0.263(2) \\
    \end{tabular}
    \caption{%
      Estimate for the position of the critical endpoints E$_{am_0}$ on the
      E$_\infty$E line (L$_+$) in the infinite volume limit.}
    \label{tab:lineee}
  \end{center}
\end{table}%
%%
%FFFFFFFFFFFFFFFFFFF Fig. 9 FFFFFFFFFFFFFFFFFFFFFFFFFFFFFFFFFFFFFFFFFF

%FFFFFFFFFFFFFFFFFFF Fig. 10 FFFFFFFFFFFFFFFFFFFFFFFFFFFFFFFFFFFFFFFFFF
\begin{figure}
  \begin{center}
    \fdfig{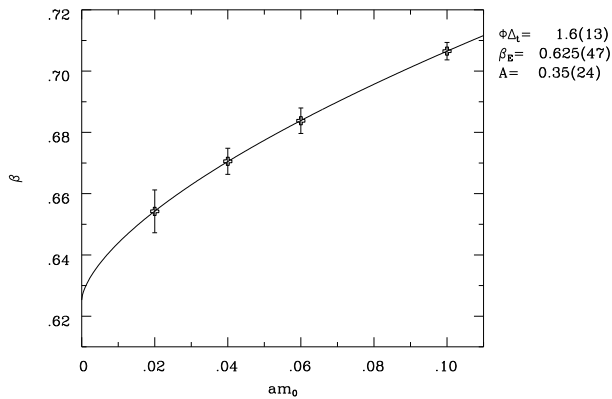}
    \caption{%
      The values of $\beta_c(am_0)$ for $am_0 \leq 0.10$ obtained from
      extrapolation into infinite volume by means of
      \protect{(\ref{BETA_PC})}. The fit shows their extrapolation into the
      chiral limit using \protect{(\ref{exbetac})}.}
    \label{fig:scalmbc}
  \end{center}
\end{figure}%
These results for $\beta_c(am_0)$ are extrapolated into the chiral limit.
The curves should approach the symmetry plane with the critical exponent
$\phi\Delta_t$:
\begin{equation}
  \label{exbetac}
  \beta_c(am_0) - \beta_c(E)
  %%  \raisebox{-.2ex}{$\stackrel{\normalsize\propto}{\sim}$}
  \propto \lambda_c(am_0) \propto (am_0)^{\phi\Delta_t}
\end{equation}
A fit of the data obtained for $am_0 \leq 0.10$ gives $\phi\Delta_t \approx
1.6$ (Fig.~\ref{fig:scalmbc}).  This value is in agreement with that
obtained by means of the relation \eqref{scal_Deltat} from the values of the exponents determined in the next section. There we find $\phi\Delta_t=1.8(1)$.
The extrapolated $\beta$ value for the point E is $\beta_{\rm E} \simeq
0.625$. Of course, with three free parameters used to fit four data points the
error is large and uncertain.

The satisfactory quality of the fit and the agreement of both methods for the
determination of $\phi\Delta_t$ indicates, that we may actually overestimate
the errors in the whole procedure. E.g. fixing $\phi\Delta_t=1.8$ in
eq.~\eqref{exbetac} reduces the error for $\beta_{\rm E}$ without reducing the
quality of the fit shown in Fig.~\ref{fig:scalmbc} and gives $\beta_{\rm
  E}=0.62(1)$.

In summary, we estimate the coordinates for the point E in infinite volume to
be:
\begin{eqnarray}
  \beta_{\rm E} &=& 0.62(3)\;, \\
  \kappa_{\rm E} &=& 0.32(2)\;.
\end{eqnarray}
The rather small improvement of the precision compared to our earlier
publication \cite{FrFr95} shows how difficult the determination of the
position of the tricritical point is, if no simulations in the symmetry plane
are possible.  Nevertheless, our present determination is much more reliable
due to the use of the scaling analysis.

%@@@@@@@@@@@@@@@@@@@@@@@@@@@@@@@@@@@@@@@@@@@@@@@@@@@@@@@@@@@@@@@@@@@@@@@
\section{Critical and tricritical exponents}
%@@@@@@@@@@@@@@@@@@@@@@@@@@@@@@@@@@@@@@@@@@@@@@@@@@@@@@@@@@@@@@@@@@@@@@@
\label{sec:kexp}%

%.......................................................................
\subsection{Exponents $\nu_t$ and $\beta_u$}
%.......................................................................
We have seen in section \ref{sec:higgsend} that the scaling behavior at the
point E$_\infty$ is mean-field-like. We now repeat the analysis by means of
Fisher zeros also for small fixed $am_0$ and determine $\tilde\nu_{am_0}$, the
value of $\tilde\nu$ at fixed $am_0$, pursuing two aims. We want to check the
universality along the E$_\infty$E line, and we want to look for a possibly
different scaling behavior in the immediate vicinity of the tricritical point
E.

To use the method of the Fisher zeros, one coupling has to be fixed. For each
$am_0$ we have fixed $\beta$ at the position of ${\rm E}_{am_0}$. At this
$\beta$ we have done simulations for different $\kappa$ and afterwards
determined the zeros in the complex $\kappa$ plane. The accuracy of the method
is limited by the uncertain precision of $\beta_c(am_0)$, as given in
table ~\ref{tab:lineee}. To estimate this dependence, we have made
measurements at three $\beta$-values for $am_0=0.02$.  Between 9600 and 64000
HMC-Trajectories have been done for each $\beta$ value at 5-7 $\kappa$-points.

%FFFFFFFFFFFFFFFFFFF Fig. 11 FFFFFFFFFFFFFFFFFFFFFFFFFFFFFFFFFFFFFFFFFF
\begin{figure}
  \fdfig{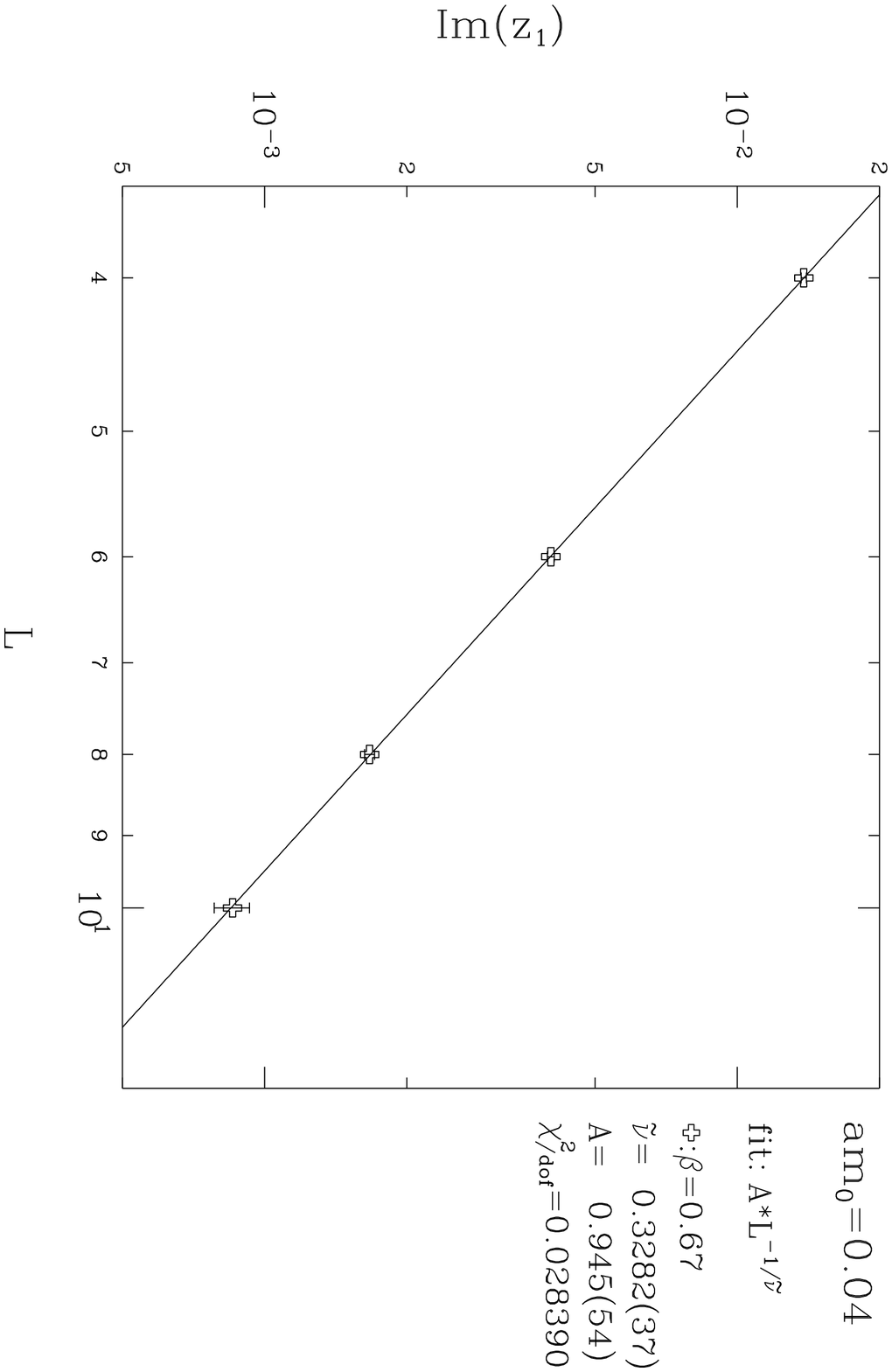}\\[5mm]
  \fdfig{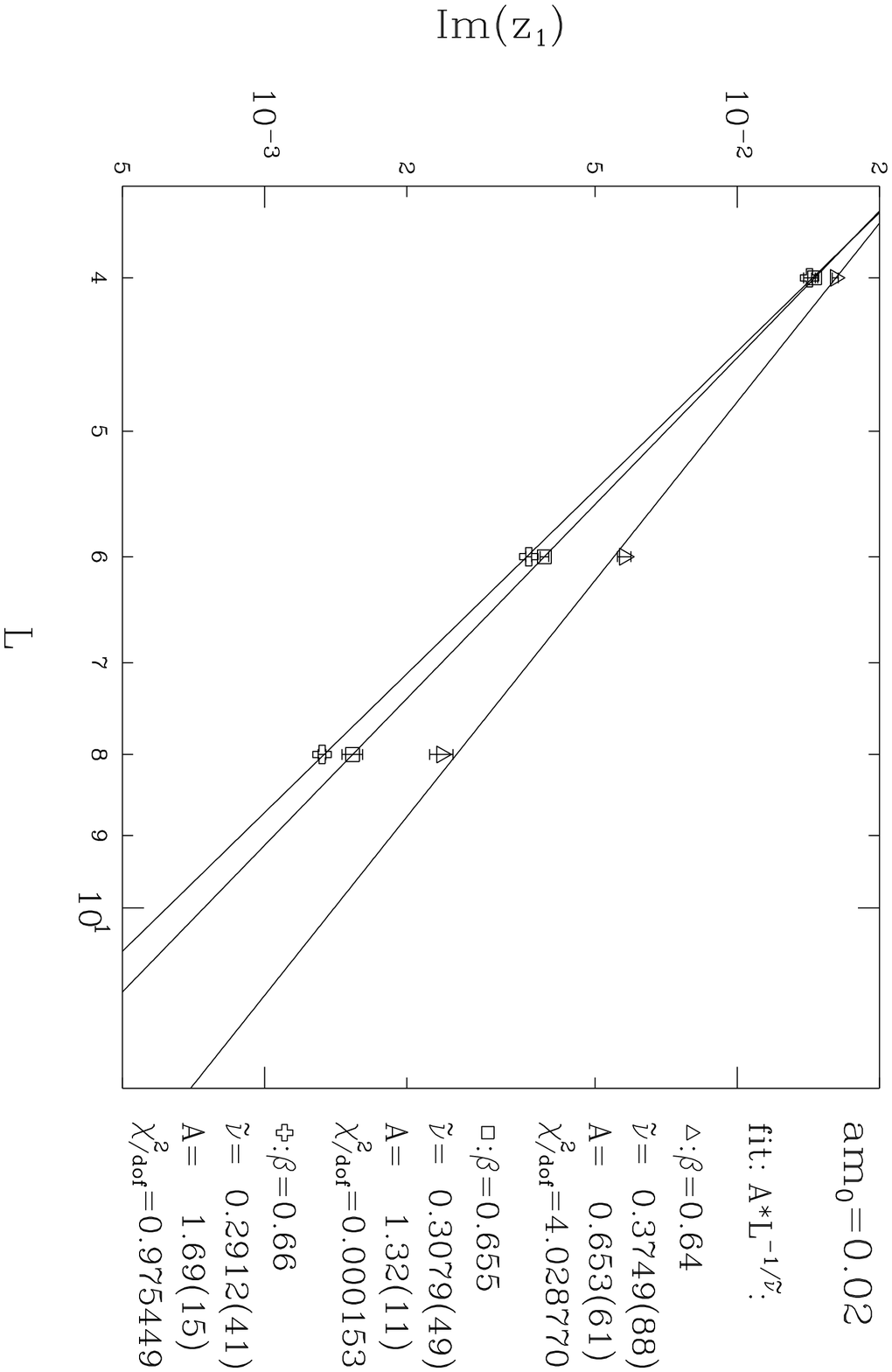}
  \caption{%
    Scaling behavior of the edge singularity: (a) for $am_0=0.04$ and
    $\beta=0.67$, (b) for $am_0=0.02$ and $\beta=0.64$, 0.655, 0.66 on
    quadratic lattices.  $\beta=0.67$ is our current best estimate for the
    position of the endpoint ${\rm E}_{am_0}$ at $am_0=0.04$. For $am_0=0.02$
    we have investigated three different $\beta$ values, $\beta=0.655$ is
    nearest to the endpoint.}
  \label{fig:hfisher}
\end{figure}%
Fig.~\ref{fig:hfisher} shows the scaling behavior of the imaginary part of the
Fisher zero in $\kappa$ for $am_0 = 0.04$ and $0.02$. In all cases, no
deviations from a linear behavior could be observed. The value $\tilde\nu_+ =
\tilde\nu_{0.04} = 0.328(4)$ for $am_0=0.04$ is in excellent agreement with
the result at $am_0=\infty$. For $am_0=0.02$ it was more difficult to find the
critical coupling $\beta_c(0.02)$. From the three measurements we estimate
$\tilde\nu_{0.02} = 0.33(4)$.

Investigation of the cumulants yields results for $\tilde\alpha_+/\tilde\nu_+$
which are consistent with the use of the Josephson relation.  This indicates
that the hyperscaling hypothesis is fulfilled. The calculation of
$\tilde\nu_+$ with use of the Josephson relation yield consistent results with
a little bit larger error.

The finite size scaling theory above the critical dimension should be applied
with care. It is possible that in spite of consistent scaling it is not the
exponent $\nu$ which is observed\footnote{We thank K. Binder for a discussion
  on this point}. In our case the values of $\tilde\nu_+$ obtained by this
method and from the scaling of the fermion mass are consistent, however.

Therefore, we interpret our results as a good confirmation of the universality
along the E$_\infty$E line. The measured values are nearly identical to those
at $am_0=\infty$. Also the logarithmic corrections seem to tend into the same
direction. Since we found similar results also for $\beta_+$ it is likely that
all exponents along the E$_\infty$E line are independent of $am_0$.

Assuming that for the small values of $am_0$, we could investigate, the
dominance region of the tricritical point is already achieved, we expect that
for $am_0 \rightarrow 0$ the exponent $\tilde\nu_+$ turns over into
$\tilde\nu_u = \nu_t$. This implies that the subsidiary exponents with the
index $u$ are identical or at least very similar to the exponents along the
E$_\infty$E line, which are the corresponding classical values. The measured
value then corresponds to $\nu_t=\frac{1}{3}$. This means, that the value is
different from the classical values of a tricritical point. The corresponding
classical prediction $\nu_t=\frac{1}{2}$ is hardly compatible with the data.

A crossover to the exponents of the tricritical point at small $am_0$ could be
expected also for the exponent $\beta_+$, which we measured along the
E$_\infty$E line (in the S$_+$ plane) in section \ref{sec:latheat}. However,
also here we could not observe any $am_0$ dependence and $\beta_+$ is
compatible with the mean field exponent of a critical line
$\beta_+=\frac{1}{2}$ down to $am_0=0.02$.  We therefore interpreted this
results as a indication, that also $\beta_u$ is close to $\frac{1}{2}$.

%.......................................................................
\subsection{Estimate of $\delta_t$ from $R_\pi$}
%.......................................................................
As suggested in \cite{KoKo93a} it is possible to determine the `magnetic'
exponent $\delta$ for fermionic theories by measuring the susceptibility ratio
$R_\pi$ for different small $am_0$ around the critical point.  We have
measured $R_\pi$ for different $\kappa$ values.  Inside the broken phase
($\kappa<\kappa_c$), we expect a curve which approaches $R_\pi=0$ for
$am_0\rightarrow0$. In the symmetric phase ($\kappa>\kappa_c$), we expect the
curve to approach $R_\pi=1$ for $am_0\rightarrow0$. At the critical point
($\kappa=\kappa_c$), the line should be horizontal for small $am_0$ and the
corresponding value of $R_\pi$ is equal to $1/\delta=1/\delta_t$.

\begin{figure}
  \begin{center}
    \fdfig{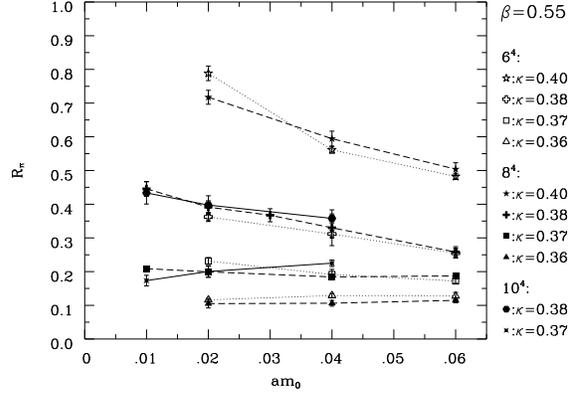}
    \caption{%
      $R_\pi$ as function of $am_0$ at $\beta=0.55$ on $6^4$, $8^4$ and
      $10^4$-lattices for different $\kappa$.}
    \label{fig:rpi055}
  \end{center}
\end{figure}

We have tried this method in our model at $\beta=0.55$ on the NE line
(Fig.~\ref{fig:rpi055}).  For $\kappa=0.37$ the curves bend downward when
approaching the chiral limit, for $\kappa=0.38$ they bend upward. This
indicates, that the critical $\kappa$ is between 0.37 and 0.38, in agreement
with 0.376(5), our estimate based on the modified gap equation \cite{FrFr95}.
The estimated horizontal line separating both phases is at $R_\pi=0.30(5)$.
This is in good agreement with the mean field value of $\delta=1/R_\pi=3$.
Both results confirm our earlier result that the phase transtion at
$\beta=0.55$ is mean-field-like \cite{FrFr95}.
%%
%FFFFFFFFFFFFFFFFFFF Fig. 12 FFFFFFFFFFFFFFFFFFFFFFFFFFFFFFFFFFFFFFFFFF

%FFFFFFFFFFFFFFFFFFF Fig. 13 FFFFFFFFFFFFFFFFFFFFFFFFFFFFFFFFFFFFFFFFFF
\begin{figure}
  \begin{center}
    \fdfig{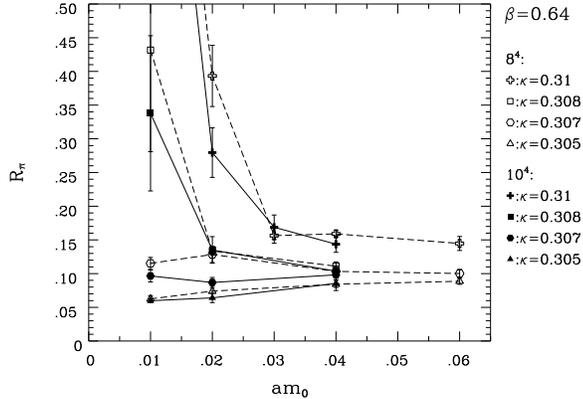}
    \caption{%
      $R_\pi$ as function of $am_0$ close to the tricritical point at
      $\beta=0.64$ on $8^4$ and $10^4$-lattices for different $\kappa$.}
    \label{fig:rpi064}
  \end{center}
\end{figure}%
Our results for $\beta=0.64$, close to the tricritical point, are shown in
Fig.~\ref{fig:rpi064}. A horizontal curve is found for $\kappa_c=0.307(2)$. It
extrapolates to $R_\pi(\kappa_c)=0.10(5)$. We estimate that
$\delta_t=10^{+10}_{-4}$ under the assumption, that the increase of the volume
does not change this picture significantly. Indeed, the data for the $8^4$ and
$10^4$ differ only very little.

To estimate the sensitivity of these results on $\beta$ we have obtained lower
statistics data also at $\beta=0.62$, our current best estimate for
$\beta_{\rm E}$. The results are very similar to those at $\beta=0.64$. This
indicates that for $\beta=0.64$ we are already in the influence region of the
tricritical point.

Also an explorative investigation with the method of the Lee-Yang
zeros in the complex $am_0$ plane, as we applied recently in the
three-dimensional model \cite{BaFr98,BaFo98}, confirms the large value
for $\delta_t$. Using the lattices $4^4$ and $6^4$ we get for
$\beta=0.64$ and $\kappa=0.307$ the estimate $\tilde\nu\simeq0.275$,
which corresponds to $\delta_t=10$.
 
%.......................................................................
\subsection{Summary of results for the tricritical exponents}
%.......................................................................
We have determined three independent exponents at the tricritical point:
\begin{eqnarray}
  \label{nut}
  \nu_t &=& 0.33(5),\\
  \label{betau}
  \beta_u &=& 0.5(2),\\
  \label{deltat}
  \delta_t &=& 10^{+10}_{-4}\,.
\end{eqnarray}
These values disagree with the classical values for tricritical points
expected in four dimensions, $\nu_t= 0.5$, $\beta_u = 1$, $\delta_t=5$.

The errors take into account only uncertainties in the measurement. We cannot
estimate possible systematic errors. In particular we have assumed that at
$am_0 = 0.02$ on the used lattices we observe the asymptotic scaling behavior
of the tricritical point.  Some support for this assumption is obtained in the
spectrum analysis in the next section. It is plausible also because typically
tricritical points have a large region of influence and the corresponding
deviations from the scaling behavior are small.
 
If one assumes the validity of the scaling laws, all further exponents are
fixed. We only could check with good precision the Josephson relation between
$\alpha_t$ and $\nu_t$. Some very crude check was possible for $\nu_u$, and
another one for $\beta_t$ is described in Sec.~\ref{sec:betat}.  We found no
indications for a violation of the scaling laws.

%@@@@@@@@@@@@@@@@@@@@@@@@@@@@@@@@@@@@@@@@@@@@@@@@@@@@@@@@@@@@@@@@@@@@@@@
\section{Spectrum}
%@@@@@@@@@@@@@@@@@@@@@@@@@@@@@@@@@@@@@@@@@@@@@@@@@@@@@@@@@@@@@@@@@@@@@@@
To analyze the possible physical content of the continuum limit taken at the
point E, it is important to study the spectrum. This investigation has the
advantage that it is relatively independent of the missing theory of
tricritical points in four dimensions. Here we give only the most important
results. A tabular overview of the measured values (which could be presented
even graphically only in part) can be obtained from the authors.

Most of the shown results have been obtained for fixed $\kappa=0.30$. The
reason for this is the observation, that the shift of the endpoint E$_{am_0}$
with $am_0$ is smaller in the $\kappa$ direction than in the $\beta$
direction. The advantage of this is, that for different $am_0$ the endpoint
E$_{am_0}$ can be approximately hit with one $\kappa$. The difference of this
chosen value of $\kappa$ from our current best estimate $\kappa_{\rm
  E}=0.32(2)$ is due to the underestimated remaining shift at the beginning of
the large scale simulation.  The value $\kappa=0.30$ corresponds the the
position of the endpoint E$_{am_0}$ at $am_0\approx 0.03$.

Some measurements with less statistics have been performed for $\beta=0.64$ and
$\kappa=0.305$. They confirm the picture presented here. In particular, they
indicate a common scaling behavior in a whole region around E, e.g.
independent of the direction of approach to this point, provided it is not
tangential to the critical lines.

%.......................................................................
\subsection{Fermion, boson and gauge-ball spectrum}
%.......................................................................
\subsubsection{Uniform behavior as a function of $am_F$}
The pictures showing the behavior of the mass of fermion and scalar boson S in
the vicinity of E have been shown already in section \ref{sec:extri}. The very
similar position of the pseudocritical area of $am_S$ and $am_F$ for different
volumes and small $am_0$ for $\kappa=0.30\simeq\kappa_{\rm E}$ suggests to
express one mass as a function of the other. 

\begin{figure}
  \begin{center}
    \fdfig{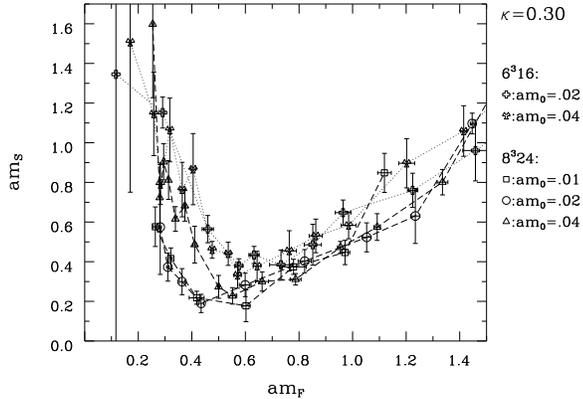}
    \caption{%
      Boson mass $am_S$ as function of the fermion mass $am_F$. The broken
      phase is to the right from the minima.}
    \label{fig:fimp1_bosc1}
  \end{center}
\end{figure}

As $am_F$ is monotonously decreasing with increasing $\kappa$ or $\beta$, and
is well measurable, we plot $am_S$ as a function of $am_F$
(fig.~\ref{fig:fimp1_bosc1}). In this figure the broken phase is to the right
and the symmetric phase to the left.  The border between both phases is at the
minimum of the boson mass. In the infinite volume and the chiral limit, the
symmetric phase would reduce in this plot to one point, $am_F = 0$. This kind
of plot as a function of $am_F$ is possible for all couplings near to the NE
line, but $am_S$ scales only at the point E.
%%
%FFFFFFFFFFFFFFFFFFF Fig. 14 FFFFFFFFFFFFFFFFFFFFFFFFFFFFFFFFFFFFFFFFFF

To extract continuum physics in this way, it is necessary to check that all
curves are close to a uniform curve, when the tricritical point is approached.
In principle, this means the fourfold limit $V\rightarrow\infty$,
$am_0\rightarrow0$, $\kappa\rightarrow\kappa_{\rm E}$ and
$\beta\rightarrow\beta_{\rm E}$.

One can see that in the broken phase, $am_S$ as a function of $am_F$ is nearly
independent of volume and $am_0$. For other $\kappa\simeq\kappa_{\rm E}$ and
$\beta\simeq\beta_{\rm E}$ nearly the same curves result. From this nontrivial
observation we conclude, that our data for fixed $am_0$ and $\kappa=0.30$
correspond approximately to a path towards E for $am_0=0$. For small $am_F$,
at the transition into the symmetric phase, the boson mass $am_S$ increases
rapidly. As expected, the minimum shifts to the left for increasing lattice
size and decreasing $am_0$, which corresponds to the approach to the critical
theory (chiral limit in the infinite volume). 

In the following, all observables are plotted as a function of $am_F$.  To
make the figures more clear, usually only the data on the $8^324$ lattice are
shown, as long as the effects of finite volume in this kind of plot are small.

\subsubsection{Scaling behavior of $am_S$}
\label{sec:scal_sf}
\begin{figure}
  \begin{center}
    \fdfig{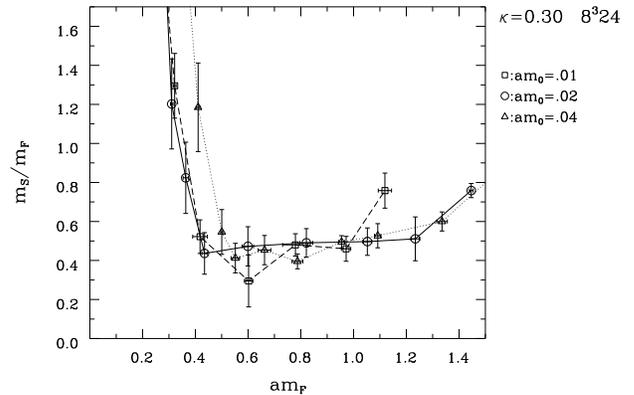}
    \caption{%
      Mass ratio $m_S/m_F$ as function of the fermion mass.}
    \label{fig:scalbos}
  \end{center}
\end{figure}

To investigate the scaling behavior of $am_S$ we look at the mass ratio
$m_S/m_F$. As shown in fig.~\ref{fig:scalbos}, this ratio is nearly constant,
and close to 0.5 in the broken phase. A small decrease of this ratio for
decreasing $am_F$ is indicated. Most probably the reason for this is, that for
nonzero $am_0$, the boson mass $am_S$ vanishes at the point E$_{am_0}$ in the
$V\rightarrow\infty$ limit, whereas $am_F$ stays finite. We have checked that
approximately the same mass ratio is obtained also on other paths into the
point E. This strongly suggests that this mass ratios is preserved also in the
continuum limit. This would mean that the scalar boson would survive with
approximately half the fermion mass.
%%
%FFFFFFFFFFFFFFFFFFF Fig. 15 FFFFFFFFFFFFFFFFFFFFFFFFFFFFFFFFFFFFFFFFFF

This similar scaling behavior means, that both observables have a common
critical point. This is our strongest argument for the coincidence of chiral
and Higgs phase transition at the point E which is thus a tricritical point.

The fact that this nice scaling behavior can be observed for all investigated
$am_0 \le 0.04$ further suggests that these values of the bare mass are inside
the dominance region of the tricritical point. In the influence region of the
E$_{\infty}$E critical line $am_F$ would stay finite whereas $am_S$ would
scale to zero.  This should at least show up in scaling deviations.

\subsubsection{Gauge balls and vector boson}
\begin{figure}
  \begin{center}
    \fdfig{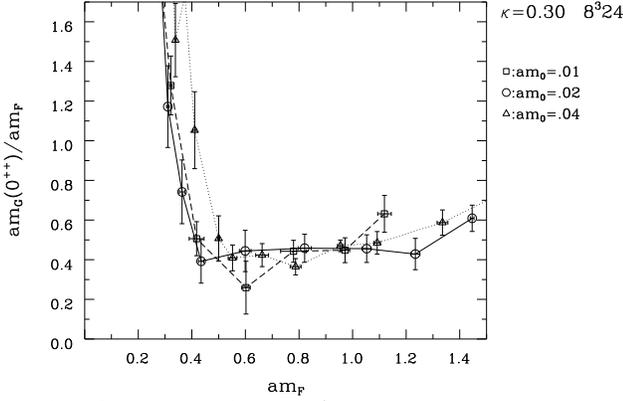}
    \caption{%
      mass ration
      $m_{\rm G}/m_F$ of the scalar gauge ball ($0^{++}$) as
      function of the fermion mass.}
    \label{fig:scalgb}
  \end{center}
\end{figure}

We also measured the mass of two gauge balls. The scalar gauge ball with the
quantum numbers $0^{++}$ (Fig.~\ref{fig:scalgb}) has nearly exactly the same
values as the scalar boson S obtained from the $\phi^\dagger-\phi$ correlation
function (\ref{ObosS}).  We have also measured the cross-correlation between
both channels at some points and found a mass in good agreement with that from
individual channels. This strongly suggests that in both channels we see one
state, which can be interpreted both as a scalar boson and gauge ball.  In the
Nambu phase, the second interpretation might be more natural as it holds in
the whole phase, including $\kappa = 0$. Nevertheless, we continue to denote
the state by S.
%%
%FFFFFFFFFFFFFFFFFFF Fig. 16 FFFFFFFFFFFFFFFFFFFFFFFFFFFFFFFFFFFFFFFFFF

We also looked at the gauge-ball channel with the quantum numbers $1^{+-}$,
but we could not observe any light particle near the tricritical point.

\begin{figure}
  \begin{center}
    \fdiifig{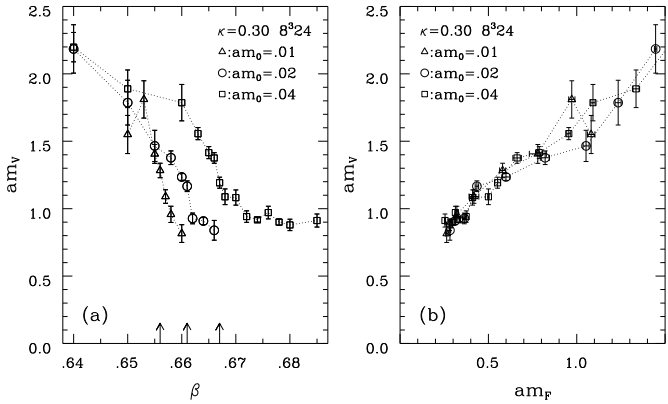}
    \caption{%
      Mass of the vector boson $am_V$ as function of (a) $\beta$ and (b)
      $am_F$ for $\kappa=0.30$ on the $8^324$ lattice. The phase transition
      (minimum of $am_S$) is marked in (a) with arrows.}
    \label{fig:bos2}
  \end{center}
\end{figure}

The mass of the vector boson $am_V$ does not scale (Fig.~\ref{fig:bos2}). The
mass decreases significantly in the symmetric phase but stays large at the
phase transition ($am_V>1$).%
%%
%FFFFFFFFFFFFFFFFFFF Fig. 17 FFFFFFFFFFFFFFFFFFFFFFFFFFFFFFFFFFFFFFFFFF

\subsubsection{Scaling of $am_F$ and $\cbcex$}
\label{sec:betat}

We have tried to estimate the tricritical exponents also from the fermion mass
$am_F$ and the chiral condensate $\cbcex$. The proper method would be to
extrapolate the data first to the infinite volume and then into the chiral
limit. As discussed in section \ref{sec:extri}, we are not able to do this.
Therefore we have investigated whether both effects can be absorbed into the
$am_0$ and volume dependence of the pseudocritical coupling $\beta_{pc}$. This
is suggested by the fact that the value of $am_S/am_F$ is nearly independent
of $am_F$ and of the used (small) $am_0$. As $am_S$ scales for $am_0\neq0$,
this should be approximately so also for $am_F$ and $\cbcex$.  ($am_S$ itself
is less suitable for such an investigation, because of the larger errors).

\begin{figure}
  \begin{center}
    \fdfig{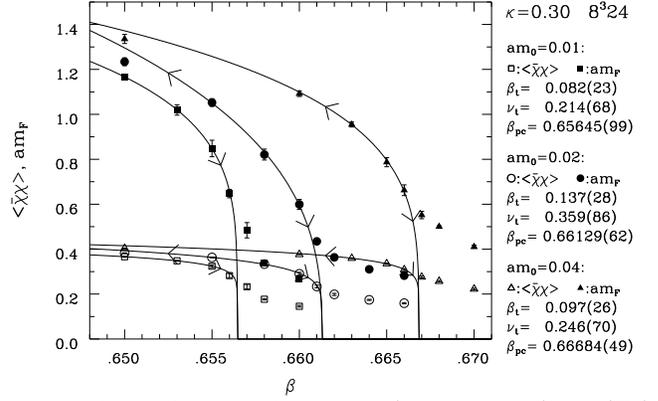}
    \caption{%
      Scaling behavior of $am_F$ (filled symbols) and $\cbcex$ (open symbols)
      as function of the coupling $\beta$ for $\kappa=0.30$ on $8^324$
      lattices. For each $am_0$ a simultaneous fit to $\cbcex$ and $am_F$ with
      one common critical coupling was done. The fit interval is marked with
      arrows. The results of the fit are given in the legend (the amplitudes
      are not shown).}
    \label{fig:pbpfimp1}
  \end{center}
\end{figure}

Correspondingly, we plot the results for the fermion mass $am_F$ and the
chiral condensate $\cbcex$ for different $am_0$ as a function of the coupling
$\beta$ (Fig.~\ref{fig:pbpfimp1}). We expect the approximate scaling behavior
\begin{eqnarray}
  am_F &=& A_F (\beta_{pc}-\beta)^{\nu_t}\\
  \cbcex &=& A_\sigma (\beta_{pc}-\beta)^{\beta_t}\;,
\end{eqnarray}
where $\beta_{pc}$ depends both on the volume and $am_0$. To get stable
results we did, for each $am_0$, a fit with a common pseudocritical coupling
$\beta_{pc}$. We choose for the fit of both observables the 3 points closest
to the critical coupling (as determined by the minimum of $am_S$) in the
broken phase. At these points the value of the condensate is still around or
above 0.3. So we end up with 6 measured values and 5 free fit parameters.
%%
%FFFFFFFFFFFFFFFFFFF Fig. 18 FFFFFFFFFFFFFFFFFFFFFFFFFFFFFFFFFFFFFFFFFF

The so estimated parameters $\beta_{pc}$, $\beta_t$ and $\nu_t$ are listed in
Fig.~\ref{fig:pbpfimp1}. The values of $\beta_{pc}$ are in nice agreement with
the minimum of the boson mass $am_S$. The results for critical exponents are
in rough agreement with the values of $\nu_t=0.33(5)$ and
$\beta_t=0.12^{+12}_{-6}$ obtained in section \ref{sec:kexp}. (The latter
value is obtained from \eqref{deltat} by means of \eqref{scal_betat} and
\eqref{scalt_joseph}.)  These results support the values of the exponents
(eq.~\eqref{nut} and \eqref{deltat}) obtained by other methods at the
tricritical point.

%.......................................................................
\subsection{Properties of $\pi$-meson}
\begin{figure}
  \begin{center}
    \fdfig{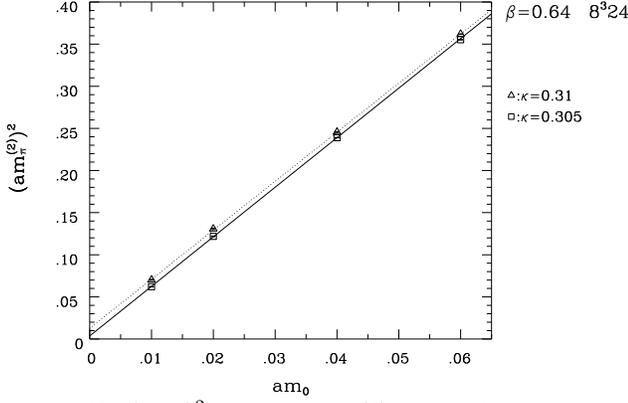}
    \caption[]{%
      $(am_\pi)^2$ from channel (2) as function of $am_0$. The PCAC relation
      \eqref{pcac_pi} corresponds to a straight line through the origin.}
    \label{fig:pi}
  \end{center}
\end{figure}

%.......................................................................
The mass of the $\pi$-meson ($0^{-+}$) can be measured very reliably. In the
broken phase we expect the validity of the PCAC relation
\begin{equation}
  \label{pcac_pi}
  (am_\pi)^2 \propto am_0\;.
\end{equation}
As shown in Fig.~\ref{fig:pi}, this relation is nearly perfectly fulfilled for
$\beta=0.64$, $\kappa=0.305$ (near to the tricritical point in the broken
phase). Thus it has the expected scaling behavior of a Goldstone boson. For
$\kappa=0.31$ (transition to the symmetric phase) an expected deviation is
observed.
%%
%FFFFFFFFFFFFFFFFFFF Fig. 19 FFFFFFFFFFFFFFFFFFFFFFFFFFFFFFFFFFFFFFFFFF

\begin{figure}
  \begin{center}
    \fdfig{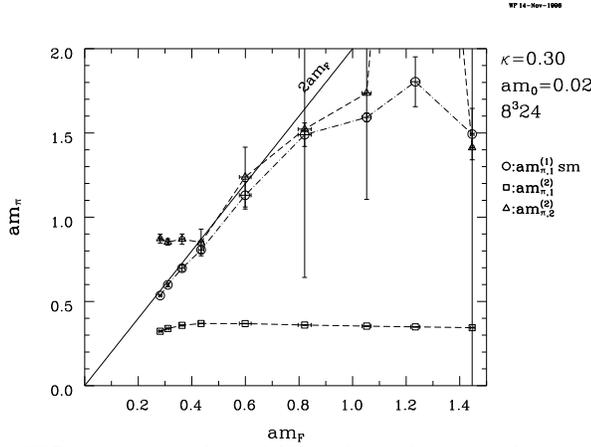}
    \caption[]{%
      Mass of $am_\pi$ from channel (1) (circles) together with mass $am_\pi$
      and first excited state of channel (2) (squares and triangles,
      respectively) as function of $am_F$. In channel (1) a simultaneous fit
      to the data with point source and smeared source has been used.}
    \label{fig:pi12}
  \end{center}
\end{figure}

To investigate the flavor symmetry restoration we compare in
Fig.~\ref{fig:pi12} the mass in the channels (1) and (2) which belongs to the
quantum numbers 0$^{-+}$ (table~I\ in \cite{FrJe95a}). These masses are
labeled $am_{\pi,1}^{(i)}$ with the corresponding channel in the exponent. In
channel (2) also the first
excited state could be measured and is labeled $am_{\pi,2}^{(2)}$.%
%%
%FFFFFFFFFFFFFFFFFFF Fig. 20 FFFFFFFFFFFFFFFFFFFFFFFFFFFFFFFFFFFFFFFFFF

A light particle can be observed only in channel (2). This is the Goldstone
boson, which scales corresponding to the PCAC relation. Its mass is different
from that in the channel (1). The first excited state in channel (2) is close
to the mass in channel (1). Both scale with approximately twice the fermion
mass. The small observed deviations for small $am_F$ are in the symmetric
phase. 

We cannot see restoration of flavor symmetry with participation of the
massless Goldstone boson. But an agreement of the massive contributions with
the corresponding quantum numbers seems to show up. This behavior of the
Goldstone boson, which is massless in the chiral limit, and its role in the
flavor symmetry restoration is not yet understood. Similar behavior was
observed e.\,g.\ in the NJL model \cite{AlGo95}.

We note that the data indicate presence of a light pseudoscalar in the
symmetric phase, as seen e.g. in Fig.~\ref{fig:pi12} at small $am_F$. In the
chiral limit the massless Goldstone particle seems to change at the phase
transition into a (bound?) state of two massless fermions $F$.

\subsubsection{Pion decay constant $f_\pi$}
\begin{figure}
  \begin{center}
    \fdfig{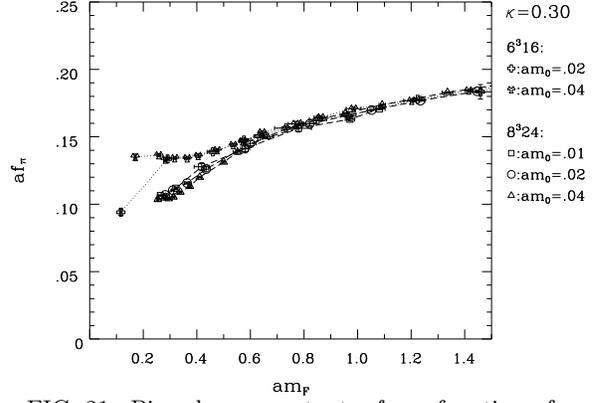}
    \caption[]{%
      Pion decay constant $af_\pi$ as function of $am_F$ for $\kappa=0.30$.}
    \label{fig:fpi030}
  \end{center}
\end{figure}

Fig.~\ref{fig:fpi030} shows the pion decay constant $af_\pi$ (eq.~\eqref{fpi})
as a function of $am_F$. The value of $af_\pi$ is nearly independent of $am_0$
and for larger $am_F$ also nearly independent of the volume. For small
$am_F$ there is a clear tendency for decrease of $af_\pi$ on larger lattices.
%%
%FFFFFFFFFFFFFFFFFFF Fig. 21 FFFFFFFFFFFFFFFFFFFFFFFFFFFFFFFFFFFFFFFFFF

%FFFFFFFFFFFFFFFFFFF Fig. 22 FFFFFFFFFFFFFFFFFFFFFFFFFFFFFFFFFFFFFFFFFF
\begin{figure}
  \begin{center}
    \fdfig{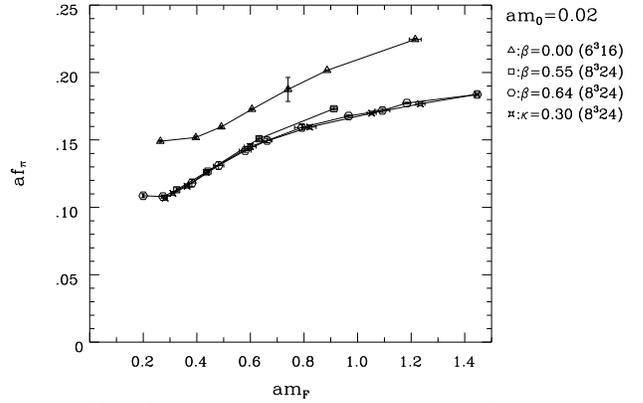}
    \caption[]{%
      Pion decay constant $af_\pi$ as function of $am_F$ for $am_0=0.02$ and
      different $\beta$ and $\kappa$, respectively, along the NE line.}
    \label{fig:fpim020}
  \end{center}
\end{figure}%
Fig.~\ref{fig:fpim020} shows a comparison of $af_\pi$ for different $\beta$
along the NE line. The smallest values for $af_\pi$ can be observed close to
the point E, but the difference between $\beta=0.55$ and $\beta=0.64$ is very
small.

If the pion decay constant $af_\pi$ scaled like a mass, its plots as a
function of $am_F$ should give a strait line through the origin.  This is not
observed, although on larger lattices there is a slight reduction observable
even in the broken phase. We do not know, if this is a hint towards a larger
finite size dependence. One may expect such a large dependence on the basis of
estimates made by means of the Schwinger-Dyson equations for the NJL model
(Fig.~33 in Ref. \cite{AlGo95}). At present we cannot exclude that
$f_\pi/m_F$ diverges in the continuum limit. This could be an indication of
the trivial continuum limit \cite{HaKo97}. In any case the largest value
$f_\pi/m_F \simeq 1/3$, we found in the broken phase around $am_F \simeq 0.4$,
is a lower bound for this ratio.

%.......................................................................
\subsection{Further mesons}
%.......................................................................
\subsubsection{$\rho$ meson mass}
The mass of the $\rho$ meson ($1^{--}$) can be measured quite well. We use a
fit to the propagator with a smeared source and sink, which suppresses excited
states quite well. We get the same mass also if we use the point source and
sink and do a fit with two states with negative (and one with positive)
parity.

\begin{figure}
  \begin{center}
    \fdiifig{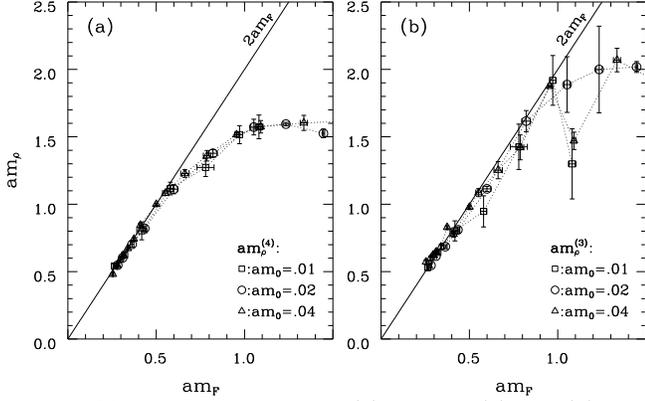}%
    \caption[xxx]{%
      $\rho$ meson mass from (a) channel (4) and (b) channel (3) for $\kappa=0.30$
      on the $8^324$ lattice. The line shows the $2am_F$ threshold.}
    \label{fig:rho}
  \end{center}
\end{figure}

The interesting feature of the $\rho$ particle is the fact, that we can
observe it in two channels. As is shown in fig.~\ref{fig:rho} the results are
nearly identical, especially near to the phase transition ($am_F\approx 0.5$).
So the flavor symmetry seems to be restored within good precision, at least
for the $\rho$.
%%
%FFFFFFFFFFFFFFFFFFF Fig. 23 FFFFFFFFFFFFFFFFFFFFFFFFFFFFFFFFFFFFFFFFFF

The values close to the phase transition are nearly exactly equal to $2am_F$
(line)\footnote{The somewhat larger values for $am_\rho$ in \cite{Fr97a} are
  due to an insufficient consideration of the excited states.} and the $\rho$
meson thus scales in excellent agreement with the fermion $F$. We cannot
distinguish, whether the $\rho$ meson is a bound state or a resonance. 

\subsubsection{$\sigma$ meson mass}
\begin{figure}
  \begin{center}
    \fdfig{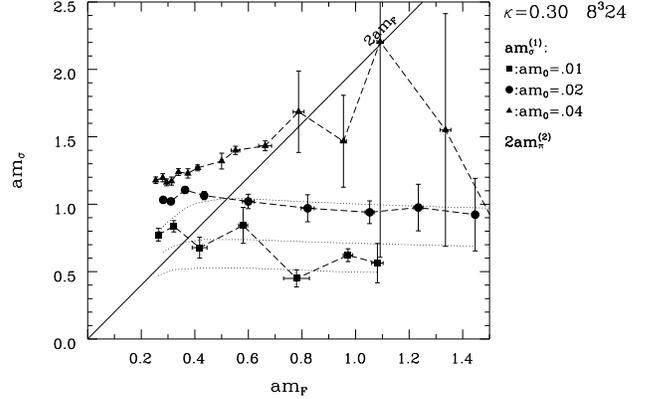}
    \caption[]{%
      $\sigma$ meson mass for channel (1) for $\kappa=0.30$ on the
      $8^324$ lattice. The full line shows the $2am_F$ threshold, the three
      dotted lines twice the $\pi$ meson mass.}
    \label{fig:sig}
  \end{center}
\end{figure}

Fig.~\ref{fig:sig} shows our measurement of the $\sigma$ meson
($0^{++}$) mass without consideration of the annihilation part. This
simplification is necessary due to computer time restrictions. It is
striking, that the $\sigma$ meson mass is especially for small $am_0$
nearly independent of $am_F$. In the broken phase it does not increase
with it. We also observe some finite size effects, with tendency of
increasing $am_\sigma$ with increasing volume. Furthermore, as
seen in Fig.~\ref{fig:sig}, the dependence on the bare mass is large.
The data therefore do not allow us to extrapolate $m_\sigma$ to the
infinite volume and chiral limits.
%%
%FFFFFFFFFFFFFFFFFFF Fig. 24 FFFFFFFFFFFFFFFFFFFFFFFFFFFFFFFFFFFFFFFFFF

It is probable that the $\sigma$ meson decays in two $\pi$ mesons. Therefore
for each $am_0$ the dotted curve shows $2am_\pi$. All measurements are within
the error bars on or above this threshold. Possibly, what we observe is not
the $\sigma$ resonance but two $\pi$ mesons. It should be noted that also in
the NJL model the $\sigma$ meson shows large finite size effects and
dependence on the bare mass\cite{AlGo95}.

The measurement of the mass of the $\sigma$ meson would have been very
interesting, as it is the most natural candidate for the Higgs boson.  Because
also the scalar boson $S$ has the same quantum numbers, we would have expected
a mixing of those two states. This does not seem to be so. Thus we have to
leave open the question, with which of the scalar particles observed on the
lattice the (composite) Higgs boson has to be identified and what mass it has.

%@@@@@@@@@@@@@@@@@@@@@@@@@@@@@@@@@@@@@@@@@@@@@@@@@@@@@@@@@@@@@@@@@@@@@@@
\section{Summary and conclusion}
%@@@@@@@@@@@@@@@@@@@@@@@@@@@@@@@@@@@@@@@@@@@@@@@@@@@@@@@@@@@@@@@@@@@@@@@

The tricritical point E in the \chup\ model turns out to be a very complex
phenomenon in four-dimensional quantum field theory, presenting numerous
challenges. No reliable way is known to study it analytically. Its numerical
investigation faces tremendous obstacles: the analysis of numerical data is
made without a plausible analytic scenario, two couplings have to be
fine-tuned, the use of dynamical fermions is necessary, and the chiral
symmetry plane is not yet accessible to simulation. The last two obstacles
make the investigation of this point substantially more difficult than the
study of tricritical points in metamagnets and other systems in statistical
mechanics.  Though statistical mechanics provides the conceptual understanding
of tricritical points which we have heavily used, there is actually no
experience with tricritical points in four dimensions to compare our results
with.

For these reasons our conclusions can be only tentative. Nevertheless,
they justify the effort\footnote{E.g. equivalent of 2 years on a CPU
  of one Gflop speed, producing about 10$^8$ HMC trajectories on
  various lattices.} we have spent: there is fair chance that the
tricritical point E exists and defines an interesting quantum field
theory in four dimensions. Though it has various features analogous to
the NJL model, the gauge field plays an important dynamical role. It
may be very complex and completely unaccessible by perturbation
theory. But it may provide a new alternative for dynamical explanation
of the masses of fundamental constituents in the standard model.

The simplest way to justify this hope is to describe our results as a
microscopic model for a strong Yukawa coupling, postulated in the standard
model e.g. for the top quark. We have done this in a separate
letter\cite{FrJe98a} \footnote{Some important technical details are given in
  the Appendix of this paper.}. In the vicinity of E, the Yukawa coupling
between F and $\pi$ emerges naturally as the Van der Waals remnant of the very
strong interactions of the \chup\ model.

The most surprising result is the clear signal for the scalar S of mass $m_S
\simeq 1/2 m_F$. It can be interpreted either as a bound state of the pair of
fundamental scalars $\phi^\dagger, \phi$, or as a gauge ball. Both channels
mix strongly. On the other hand, the would-be Higgs boson, the fundamental
fermion-antifermion bound state $\sigma = \overline{\chi} \chi$, is elusive.
It does not show scaling properties allowing an extrapolation to the continuum
limit. It does not mix appreciably with S.

The most perplexing result is the value of the tricritical exponent
$\nu_t \simeq 1/3$, which is a nonclassical value. Could it mean that
the continuum theory is nontrivial? Standard lore in statistical
mechanics is that in and above three dimensions tricritical points are
classical \cite{LaSa84}. However, this is based on the experience with
spin systems and scalar fields. Strongly coupled gauge theory with
fermions might be different. Thus point E is a challenge also for
statistical mechanics.

We feel that our aim to understand the tricritical point in the \chup\ 
model was a little bit ahead of time. Though we used the most advanced
methods, they were not powerful enough. The computational resources
should be also larger by 1-2 orders of magnitude. Algorithms for
simulation in the chiral symmetry limit are needed. Last but not
least, the interest in a search of strongly coupled theories beyond
the standard model should be much higher than the current widerspread
beliefs.

\acknowledgments We thank K.~Binder, V.~Dohm, M.~G\"ockeler, D.~Kominis,
K.-I.~Kondo, M.~Lindner, M.-P.~Lombardo, and E.~Seiler for discussions. The
computations have been performed on the Fujitsu S600, VPP500, and VPP300 at
RWTH Aachen, and on the CRAY-YMP and T90 of HLRZ J\"ulich. The work was
supported by DFG.

\appendix
\section{Details about the used operators}
\subsection{Smeared sources for meson propagators}
\label{app:smmes}
To reduce the contribution of excited states in the meson propagators we have
implemented gauge invariant smeared sources \cite{Gu90}. We are not aware of any
such an implementation for staggered fermions. It requires that the
even-odd separation is preserved. Therefore we have transported the source
with two link term to the next to nearest neighbour. This reads as
\begin{eqnarray}
  &&\phi(\vec{x},t) \rightarrow \phi^\prime (\vec{x},t) =
  \frac{1}{1-6\alpha} \nonumber\\
  && \hspace{3mm} \Bigg\{ \phi(\vec{x},t) + \alpha \sum_{i=1}^3 \big[
  U_i(\vec{x},t)U_i(\vec{x}+\vec{e_i},t)\phi(\vec{x}+2\vec{e_i},t) \nonumber\\
  && \hspace{10mm}
  +U_i^*(\vec{x}-\vec{e_i},t)U_i^*(\vec{x}-2\vec{e_i},t)\phi(\vec{x}-2\vec{e_i},t)
  \big] \Bigg\}.
\end{eqnarray}
We have chosen values $\alpha=0.01$ and $\alpha=0.02$ and 20 smearing
iterations. The resulting smearing radius was $R\simeq 0.89$ and $R\simeq
1.47$, respectively. The latter gave the better results. Compared to QCD these
radii may seem very small, which might be due to larger masses in our case,
however.

\subsection{Effective Yukawa coupling $y_{\text{R}}$}
\label{app:yr}
We have done  measurements of $y_R$ in the momentum space, similar to
\cite{GoHo95}. The meson-fermion three point function is
\begin{eqnarray}
  G_3^{(\alpha)}(p,q)&=&\frac{1}{T}\sum_{t_1,t_2}e^{-ip_4t_1+iq_4t_2}
  \sum_{\underline{x}_1,\underline{x}_2}
  e^{-i\vec{p}\underline{x}_1+i\vec{q}\underline{x}_2} \nonumber\\
  &&\Bigg\langle \phi(\underline{x}_1,t_1)\overline{\chi}(\underline{x}_1,t_1)
    \sum_{y_4}e^{i(p_4-q_4)y_4} \nonumber\\
  &&\quad M_\alpha(\vec{p}-\vec{q},y_4)
      \phi^\dagger(\underline{x}_2,t_2)\chi(\underline{x}_2,t_2) \Bigg\rangle
  \, ,
\end{eqnarray}
with
\begin{equation}
  M_\alpha(\vec{p}_0,y_4)=\sum_{\vec{y}} e^{i\vec{p}_0\vec{y}}
  \varphi_\alpha(\vec{y}) \overline{\chi}(\vec{y},y_4)\chi(\vec{y},y_4)\, ,
\end{equation}
and $\varphi_\sigma(\vec{y})=1$ for the $\sigma$ meson (scalar), and
$\varphi_\pi(\vec{y}) = \eta_4(\vec{y}) = (-1)^{y_1+y_2+y_3}$ for the $\pi$
meson (pseudo-scalar).  The underscore of $\underline{x}_1$ and
$\underline{x}_2$ indicates that the spatial part of the vector has even
coordinates. The corresponding sums run over the $2^3$ cubes of the spatial
lattice. The measurements were done for the momenta
\begin{eqnarray}
  &&p_4= \left\{
  \begin{tabular}[c]{ll}
    $q_4$ & for the $\sigma$ meson ,\\
    $q_4+\pi$ & for the $\pi$ meson ,
  \end{tabular}\right.
  \hspace{1cm} q_4 = \pm \frac{\pi}{T} \\
  \label{mom_yr}
  && \hspace{7mm}(\vec{p},\vec{q}) =
  (\vec{0},\vec{0}),\; (\frac{2\pi}{L}\vec{e_1},\vec{0})\,,
  (\frac{2\pi}{L}\vec{e_1},\frac{2\pi}{L}\vec{e_1})\,.
\end{eqnarray}
This choice of $p_4$ guarantees, that only states of the right parity
contribute. For $q_4$ only the smallest possible values are considered and the
results are averaged over both signs of $q_4$. The different spatial momenta
are evaluated separately. The spatial momentum of the meson is
$\vec{p}_0=\vec{p}-\vec{q}$.

For the implementation we  neglect the disconnected parts
and write by means of the fermion matrix $M_{yx}$
\begin{eqnarray}
  \label{mesg3}
  G_3^{(\alpha)}(p,q) &=& -\frac{1}{T} \left\langle \sum_{y_4,\vec{y}}
    e^{i(p-q)y} \varphi_\alpha(\vec{y})\varepsilon(y) \right. \nonumber\\
  && \left\{ \sum_{t_1,\underline{x}_1} M^{-1}_{y\,x_1} \phi(x_1)e^{-ipx_1}
    \right\} \nonumber\\
  && \left.\left\{ \sum_{t_2,\underline{x}_2} M^{-1}_{y\,x_2} \phi(x_2)
      e^{-i(q+\left(\vec{0},\pi)\right)x_2} \right\}^*\: \right\rangle,
\end{eqnarray}
where the four-vectors $x_1=(\underline{x}_1,t_1)$, $y=(\vec{y},y_4)$, \ldots\ 
have been introduced. In this notation it is well noticeable, that for our 6
different momentum pairs $(p,q)$ for both $G_3$ all together 8 matrix
inversions are needed. Due to the source on the whole lattice the signal is
very good and we have done a measurement only on each $8^{\rm th}$
configuration.

The effective Yukawa coupling $y_{\text{R}}^{(\alpha)}$ is now
obtained from the comparison of the Monte-Carlo data for the
three-point function, with that obtained in the tree level
approximation of an effective lattice action, which describes the
interaction of the staggered fermion fields $F$, $\overline{F}$ and a
(pseudo-)scalar field $\Phi$ with a coupling term of the form
\begin{eqnarray}
  -y_{\text{R}}^{(\sigma)} \sum_x \Phi(x)\overline{F}(x)F(x) &\quad& \text{(scalar)}\;,\\
  -y_{\text{R}}^{(\pi)} \sum_x \varepsilon(x)
  \Phi(x)\overline{F}(x)F(x) &\quad& \text{(pseudoscalar)}.
\end{eqnarray}
The connected part of the three-point function is
\begin{eqnarray}
  \label{defyr}
  &&G_3^{(\alpha)}(p,q) = - \frac{y_{\text{R}}^{(\alpha)}}{T}\frac{V}{8}
  \sum_{\vec{\omega}} \bar\theta_\alpha(\vec{\omega})
  \tilde{G}_\alpha(p-q) \nonumber\\
  &&\hspace{4mm}\left\{ \sum_{\underline{x}_2,t_2}
    \eta_4(\vec{\omega})^{t_2} e^{iq(x_2+\vec{\omega})}
    G_F(x_2+\vec{\omega},0)_{\vec{q}}\right\} \nonumber\\
  &&\hspace{4mm}\left\{ \sum_{\underline{x}_1,t_1}
    \eta_4(\vec{\omega})^{t_1}
    e^{i\left(p+(\vec{0},\pi)\right)(x_1+\vec{\omega})}
      G_F(x_1+\vec{\omega},0)_{\vec{p}} \right\}^*
\end{eqnarray}
with
\begin{equation}
  \bar\theta_\sigma(\vec\omega) = \eta_4(\vec\omega)\;,\qquad
  \bar\theta_\pi(\vec\omega) = 1\,.
\end{equation}
Here $\vec{\omega}$ runs over corners of the elementary three-dimensional
cube. In the tree-level approximation, $\tilde{G}_\alpha$ and $G_F$ are the
free propagators for the meson and the fermion, the tilde indicating the
Fourier transformation of the meson propagator. These propagators are replaced
by the full propagators from the simulation and the wavefunction
renormalization constants $Z_F$ and $Z_\alpha$ are included:
\begin{eqnarray}
  \tilde{G}_\alpha(p_0) &=& \sqrt{Z_\alpha(\vec{p}_0)} \theta_\alpha
  \sum_{x} e^{ip_0x} \nonumber\\
  &&\hspace{7mm}\left\langle \varphi(\vec{x}) \overline{\chi}(x)\chi(x)\;
  \varphi(\vec{0} )\overline{\chi}(0)\chi(0) \right\rangle_c \\
  G_F(x,0)_{\vec{p}} &=&  \sqrt{Z_F(\vec{p})} \left\langle \overline{\chi}(x)\chi(0)
  \right\rangle\, ,  
\end{eqnarray}
with $\theta_\sigma=-1$ and $\theta_\pi=1$ to correct for the negative sign
of the $\sigma$ propagator.

After these replacements and identifications of $G_3^{(\alpha)}(p,q)$
with the measured values of expression (\ref{mesg3}) we obtain
$y_{\text{R}}^{(\alpha)}(p,q)$ from eq.  (\ref{defyr}). $y_{\text{R}}$
should be real and only slightly dependent on the momenta. For the
determination of $Z(\vec{p})$ we have measured the corresponding
propagators at momentum $\vec{p}$ and then performed a fit with the
free propagator.

Using these definitions we have measured the real and imaginary part of
$y_{\text{R}}$. The implemented momenta combinations have been
numbered corresponding to equation (\ref{mom_yr}) from 1 to 3.
%%
%FFFFFFFFFFFFFFFFFFF Fig. 25 FFFFFFFFFFFFFFFFFFFFFFFFFFFFFFFFFFFFFFFFFF
\begin{figure}
  \begin{center}
    %% \fdfig{fig/h4k030m020CK.fimp1_yrsig.eps}\\[3mm]
    \fdfig{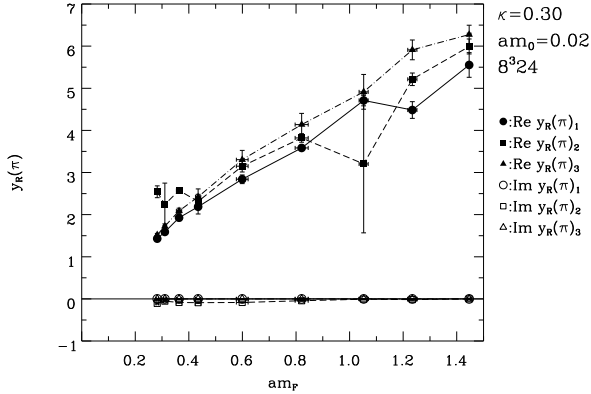}
    \caption{
      Results for the real and imaginary part of $y_{\text{R}}$ of the $\pi$
      meson for different momenta at $\kappa=0.30$ and $am_0=0.02$ on the
      $8^324$ lattice as a function of $am_F$.}
    \label{fig:test_yr}
  \end{center}
\end{figure}%

For the effective coupling of the $\pi$ meson we get a consistent picture
(Fig.~\ref{fig:test_yr}). The imaginary part of $y_{\text{R}}$ is very small,
and for different momenta we get approximately the same values. Because of
this agreement we restrict ourselves to the evaluation of the data with
vanishing momentum (No.~1).

For the $\sigma$ meson we failed to get a reliable $y_{\text{R}}$. The
imaginary part is not really small and the real part is only for the momentum
combinations 1 and 2 (at least one fermion momentum vanishes) approximately
equal. The inconsistencies are larger in the broken phase. The problems might
be related to the neglected disconnected parts but also to the fact, that the
$\sigma$ meson is probably only a broad resonance. In this sense this
measurement shows again the problems we met already in the measurement of the
$\sigma$ meson mass.

%%%%%%%%%%%%%%%%%%%%%%%%%%%%%%%%%%%%%%%%%%%%%%%%%%%%%%%%%%%%%%%%%%%%%%

\bibliographystyle{wunsnot}   % wunsnot style (unsorted numbers, no article titles)
%%\bibliographystyle{wunstit}   % wunstit style (unsorted numbers, article titles)
%\bibliographystyle{wcitenot}  % wcitenot style (sorted cite$, no article titles)
%%\bibliographystyle{wcitetit}  % wcitetit style (sorted cite$, article titles)
%%\bibliographystyle{wabst}     % wabst style (unsorted cite$, article
%%titles, abstracts, ...)
% \bibliography{jourabbr,our-papers,u1,referen,gauge,yukawa}
% begin_of_bibliography

% end_of_bibliography

\end{document}